\documentclass[12pt]{book}

\usepackage{graphicx}
\usepackage{amssymb,amsmath,amsfonts,amsthm}
\usepackage{multirow}
\usepackage{bm}
\usepackage{anysize}
\usepackage{array}
\usepackage{epstopdf}
\usepackage{eurosym}
\usepackage{color}
\usepackage{etoolbox}
\usepackage{makeidx}
\usepackage{ifthen}
\usepackage[hidelinks]{hyperref}
\usepackage{longtable}
\usepackage{titlesec} 
\usepackage{fancyhdr}
\usepackage[noadjust]{cite}
\usepackage{lscape}
\usepackage{titlesec}
\usepackage{pdfpages}
\usepackage{acronym}

\usepackage[english,greek,spanish]{babel}
\usepackage{csquotes}

\definecolor{NavyBlue}{cmyk}{0.94,0.54,0,0} % PANTONE 293

\marginsize{2.8cm}{2.8cm}{2.8cm}{2.8cm}

\acrodef{vsc}[VSC]{Voltage Source Converter}
\acrodef{adn}[ADN]{Active Distribution Network}
\acrodef{ltc}[LTC]{Load Tap Changer}
\acrodef{im}[IM]{Induction Motor}
\acrodef{dg}[DG]{Distributed Generation}
\acrodef{hvdc}[HVDC]{High Voltage Direct Current}
\acrodef{cc}[CC]{Current Controller}
\acrodef{val}[VAL]{Virtual Admittance Loop}
\acrodef{dval}[DVAL]{Dynamic Virtual Admittance Loop}
\acrodef{qval}[QVAL]{Quasi-stationary Virtual Admittance Loop}
\acrodef{pll}[PLL]{Phase-Locked Loop}
\acrodef{cf}[CF]{Complex Frequency}
\acrodef{gfm}[GFM]{Grid-Forming}
\acrodef{gfl}[GFL]{Grid-Following}
\acrodef{vsm}[VSM]{Virtual Synchronous Machine}
\acrodef{snb}[SNB]{Saddle Node Bifurcation}
\acrodef{hb}[HB]{Hopf Bifurcation}
\acrodef{lib}[LIB]{Limit Induced Bifurcation}
\acrodef{sib}[SIB]{Singular Induced Bifurcation}
\acrodef{im}[IM]{Induction Motor}
\acrodef{zip}[ZIP]{constant Impedance, constant Current, constant Power}

\makeatletter
\renewcommand\part{
  \if@openright
    \cleardoublepage
  \else
    \clearpage
  \fi
  \thispagestyle{empty}
  \if@twocolumn
    \onecolumn
    \@tempswatrue
  \else
    \@tempswafalse
  \fi
  \null\vfil
  \secdef\@part\@spart}
\makeatother

\begin{document}
\selectlanguage{english}
\frontmatter

\begin{titlepage}
\centering
\color{NavyBlue}

\centerline{\includegraphics[height=3cm]{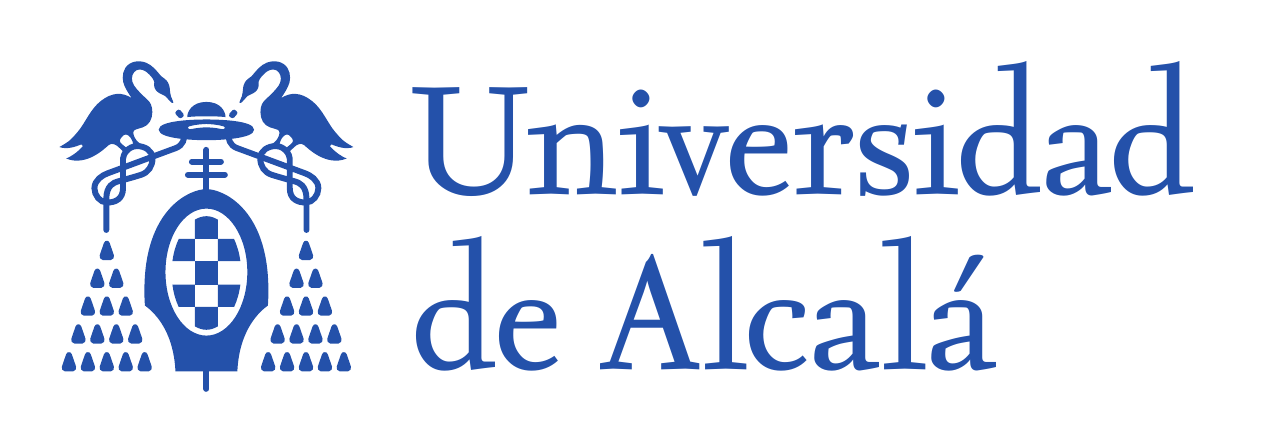}}

\begin{LARGE}
PhD. Program in Electronics: Advanced Electronic Systems. Intelligent Systems \par
\vspace{0.8in}

\textbf{Voltage Stability and Control of Electrical Distribution Systems with High Penetration of Power Electronic Converters}\par

\vspace{0.8in}
PhD. Thesis Presented by \\

\textbf{Dionysios Moutevelis} \par
\vspace{0.6in}

\par
\vspace{0.5in}
\vfill
\vspace{0.3in}

\color{black}
2024

\end{LARGE}
\par
\end{titlepage}
\thispagestyle{empty}
	
\begin{titlepage}
\centering

\centerline{\includegraphics[height=3cm]{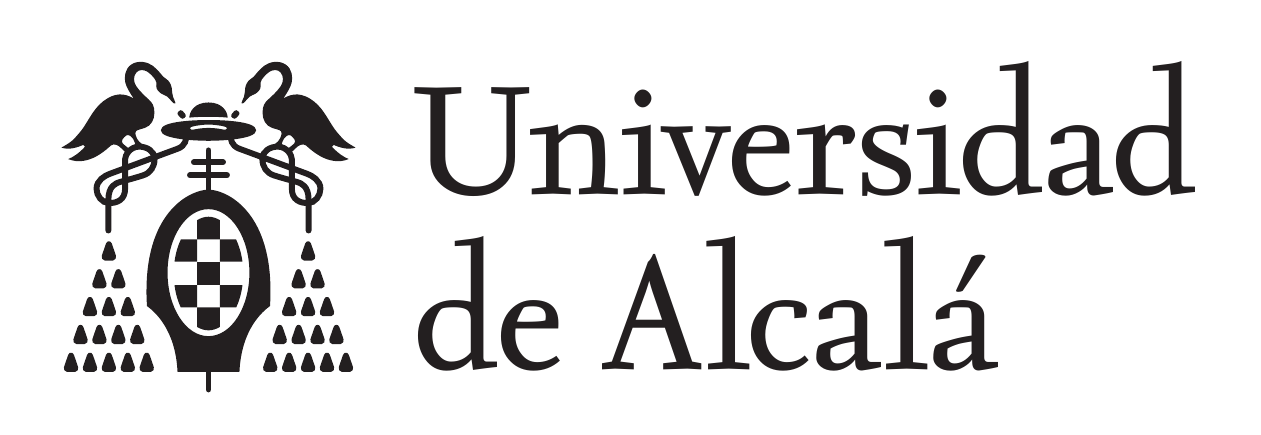}}

\vspace*{0.5in}
\begin{Large}\bfseries
Voltage Stability and Control of Electrical Distribution Systems with High Penetration of Power Electronic Converters\par
\end{Large}
\vspace{0.5in}
by \\
Dionysios Moutevelis \\
\vspace{0.5in}
supervised by \\
Dr. Javier Rold\'{a}n P\'{e}rez \\
Dr. Milan Prodanovic \\

\vspace{1in}
a dissertation submitted to University of Alcal\'{a} \\
in partial fulfilment of the requirements for the 
\\
degree of doctor of philosophy
\par
\vspace{0.5in}
\vfill
\vspace{0.3in}
Alcalá de Henares, 2024
\par
\end{titlepage}

\thispagestyle{empty}

\cleardoublepage

%\includepdf[pages=1]{ImagenesCover/VB_directores_Alberto_Rodriguez_Cabero.pdf}

\thispagestyle{empty}

\cleardoublepage

%\includepdf[pages=1]{ImagenesCover/VB_tutor_Alberto_Rodriguez_Cabero.pdf}

\thispagestyle{empty}

\cleardoublepage

%\includepdf[pages=1]{ImagenesCover/VB_coordinador_Alberto_Rodriguez_Cabero.pdf}

\thispagestyle{empty}

\cleardoublepage

\thispagestyle{empty}

\vspace*{+4cm}

\selectlanguage{greek}
\begin{displayquote}
\raggedleft \em  Στους γονείς μου, Ανδριάνα και Στέλιο \\
\end{displayquote}
\selectlanguage{english}

\vspace{+6cm}

% \begin{quotation}

% \noindent I am enough of an artist to draw freely upon my imagination. Imagination is more important than knowledge. Knowledge is limited. Imagination encircles the world.
% \medskip
% \raggedleft
% \em Albert Einstein

% ``There is nothing outside of yourself that can ever enable you to get better, stronger, richer, quicker, or smarter. Everything is within. Everything exists. Seek nothing outside of yourself.''

% \medskip
% \raggedleft
% \em The Book of Five Rings, Miyamoto Musashi

% \end{quotation}

\vspace*{\fill}

\clearpage
\thispagestyle{empty}
	\chapter*{Acknowledgements}
\thispagestyle{empty}

In every important milestone, it is important to acknowledge all the people that contributed to its achievement.
%Without them, the completion of this thesis would not have been possible.
%
First and foremost, I wish to thank my supervisor and mentor Dr. Javier Rold\'{a}n-P\'{e}rez for generously sharing his knowledge, for his continuous motivation to be better and for his support in all matters, research related or otherwise.
%, throughout the years of my PhD studies.
His creativity and persistence, while always maintaining a positive attitude, are truly inspirational.
I would also like to thank my co-supervisor, Dr. Milan Prodanovic, for giving me the opportunity to come to Spain and for providing me with advice and guidance during the development of this thesis.

I am grateful to Prof. Federico Milano for his excellent hospitality and productive collaboration during my stay in University College Dublin, Ireland as a visiting researcher. His resolute quest for knowledge motivated me to be more determined and daring in my efforts.
I would also like to thank Prof. Emilio Bueno from Alcal\'{a} de Henares University for his time and valuable feedback.

In addition, I would like to thank the people from IMDEA Energy Institute that contributed in creating a pleasant working environment. Special thanks go to Diana, Pablo and Njegos for sharing with me their PhD journey and for being outstanding office mates. Also, to our lab engineers, Alex and Josemi for always providing technical support and for all the pleasant time we spent together.

I would like to acknowledge the Community of Madrid for providing the funding for the development of this thesis via the Research Project PROMINT-CM under Grant P2018/EMT436.

Finally, I wish to express my gratitude to the people closest to me. To all my friends in Athens for receiving me every time like I never left. To my family, my parents Andriana and Stelios and my brother Nikos, for their unconditional love and support. To all the dear companions in Madrid, especially Dimitris and Alexandra, for turning these years into a life-changing experience.

\clearpage
\thispagestyle{empty}
	\chapter*{Summary}

Power systems are currently undergoing a rapid paradigm change in their operation. Centralised energy production is being replaced by a number of \ac{dg} units that are placed at different locations and voltage levels in power networks. These distributed units are mostly based on renewable energy technologies, like wind turbines and photovoltaic cells and are commonly interfaced to the grid via power electronic converters. These sources reduce energy system dependency on conventional generation units based on fossil fuels. At the same time, this shift introduces technical challenges for the safe and reliable operation of electricity networks since \ac{dg} sources do not inherently provide the grid regulation services of conventional, centralised generation units. Moreover, the increased penetration of renewable energy sources and their converter-based interfaces is creating voltage deviation and voltage stability issues in distribution networks. These issues range from overvoltages during hours of peak renewable generation, reverse power flows and sudden voltage drops due to the variable nature of renewable energy production. All of the above jeopardise the reliable operation of the distribution networks that were not originally designed to accommodate for these effects. The objective of this thesis is to propose novel techniques for the accurate assessment of the \ac{dg} impact on voltage stability in distribution networks and investigate how the control capabilities of converter-based interfaces of \ac{dg} units can be harnessed to improve stability margins and overall system robustness and performance.
%
% Additionally, the control design of these distributed sources presents high complexity due to their diverse dynamic operation and their significant number.
% %
% The location of many renewable energy sources is located close to energy consumption centers. For this reason, one subcategory of the electricity network that is greatly affected by the introduction of \ac{dg} are distribution networks. These networks operate in low voltage and are used to distribute electrical energy to consumers. They are characterised by less controllability from the grid operator viewpoint, are highly resistive and in the past were considered widely as passive grids. 

Stability analysis for distribution networks with increased penetration of power converters is extensively performed to distinguish safe operating regions. However, the small-signal techniques that are typically used are not suitable for those cases when system parameters deviate significantly from their nominal values. In addition, the effects on stability of traditional voltage regulating devices, like transformers equipped with \acp{ltc}, and of the converter capacity limits, are scantly considered. This thesis proposes bifurcation analysis as an effective tool for the stability assessment of active distribution networks with large deviations from the system nominal parameter values. The approach considers the full dynamics of the participating converters and the network and incorporates non-smooth limitations by means of smooth approximations. The analysis predicts accurately the stability margins of the system and offers insight on the converter control parameter design.

Addressing voltage deviations in distribution networks via converter control is a well-established research topic. Reactive power support during undervoltages in inductive networks and active power curtailment during overvoltages in resistive networks are standard techniques for restoring bus voltage magnitudes within limits. However, a holistic approach that is suitable for both under- and overvoltages and for networks with a wide range of resistance-to-reactance ratios has not been found in the literature. This thesis proposes a control solution that effectively mitigates voltage deviations by affecting both active and reactive power outputs of the converter in a coupled way. The controller is based on the concept of virtual admittance and is implemented across different control layers, both in the converter and the distribution network levels. Implementation aspects, dynamic performance analysis and parameter control design are thoroughly addressed.

In power system analysis and control applications, voltage and frequency are typically treated independently. However, in modern power systems, the integration of converter-based devices has increased the coupling between the two system variables. For this reason, there is an increased demand for analytical tools that quantify the impact of power converters on both the voltage and frequency of such networks. In this thesis, a taxonomy of power converter
control schemes based on the concept of complex frequency is proposed. This approach captures transient variations of both voltage and frequency in a single complex quantity. The proposed framework allows the decoupling of the effects on voltage and frequency of different participating controllers in complex control structures. It also allows the identification of critical parameters, facilitating the control design.

In this thesis, different contributions regarding voltage stability and control in active distribution systems are presented. For each one, a comprehensive review of the state of the art is presented. The study thoroughly investigates the impact of power converter interfaced \ac{dg} on the voltage stability of modern distribution systems and also proposes novel control solutions for the identified issues. Furthermore, the research advances beyond the theoretical developments by experimentally validating the proposed stability analyses and by implementing and experimentally evaluating the control schemes within the Smart Energy Integration Lab (SEIL) at IMDEA Energy Institute. Finally, the thesis concludes by summarizing its findings and contributions as well as offering guidelines for future research directions.
	\selectlanguage{spanish}

\chapter*{Resumen}

Los sistemas de energía están experimentando actualmente un rápido cambio de paradigma en su funcionamiento. La producción centralizada de energía está siendo reemplazada por numerosas unidades de generación distribuida (GD) que están dispersas por toda la red. Estas unidades distribuidas suelen estar conectadas a la red a través de convertidores electrónicos de potencia y se basan en tecnologías de energía renovable, como aerogeneradores y paneles fotovoltaicos. Por esta razón, reducen la dependencia del sistema de las unidades convencionales de generación de energía que se basan en combustibles fósiles. Sin embargo, al mismo tiempo, este cambio introduce desafíos técnicos para la operación segura y confiable de las redes eléctricas, ya que las fuentes de GD no proporcionan inherentemente los servicios de regulación de la red que ofrecen las unidades convencionales de generación centralizada. En particular, la creciente penetración de fuentes de energía renovable y sus interfaces basadas en convertidores está generando problemas de desviación de tensión y estabilidad de tensión en las redes de distribución. Estos problemas van desde sobretensiones durante las horas de máxima generación renovable, flujos de potencia inversa y caídas repentinas de tensión debido a la naturaleza intermitente de la producción de energía renovable. Todo lo anterior pone en peligro la operación confiable de las redes de distribución, que originalmente no fueron diseñadas para acomodar estos efectos. El objetivo de esta tesis es proponer nuevas técnicas para la evaluación precisa del impacto de la GD en la estabilidad del tensión de las redes de distribución e investigar cómo se pueden aprovechar las capacidades de control de las interfaces basadas en convertidores de las unidades de GD para mejorar los márgenes de estabilidad y el rendimiento general del sistema.

El análisis de estabilidad en sistemas de distribución con una alta penetración de convertidores de energía se realiza ampliamente para distinguir las regiones de operación segura. Sin embargo, las técnicas comúnmente empleadas de pequeña señal no son adecuadas para casos en los que los parámetros del sistema se desvían significativamente de sus valores nominales. Además, los efectos en la estabilidad de los dispositivos reguladores de tensión tradicionales, como los transformadores equipados con reguladores de tensión tap cambiante, y de los límites de capacidad de los convertidores, se consideran escasamente. Esta tesis propone el análisis de bifurcación como una herramienta efectiva para la evaluación de la estabilidad de las redes de distribución activa con grandes desviaciones de los valores nominales del sistema. El enfoque considera la dinámica completa de los convertidores participantes y la red, e incorpora limitaciones no suaves mediante aproximaciones suaves. El análisis predice con precisión los márgenes de estabilidad del sistema y ofrece información sobre el diseño de parámetros de control de los convertidores.

Abordar las desviaciones de tensión en las redes de distribución a través del control de convertidores es un tema ampliamente investigado. El soporte de potencia reactiva durante subtensiónes en redes inductivas y la limitación de potencia activa durante sobretensiónes en redes resistivas son técnicas estándar para devolver las magnitudes de tensión del bus dentro de los límites. Sin embargo, falta en la literatura un enfoque integral que sea adecuado tanto para subtensiónes como para sobretensiónes y para redes con una amplia gama de relaciones resistencia-reactancia. Esta tesis propone una solución de control que mitiga de manera efectiva las desviaciones de tensión afectando tanto la potencia activa como la reactiva de salida del convertidor de manera acoplada. El controlador se basa en el concepto de admitancia virtual y se implementa en diferentes capas de control, tanto locales como centralizadas. Se abordan en detalle aspectos de implementación, análisis de rendimiento dinámico y diseño de control de parámetros.

Típicamente, en las aplicaciones de análisis y control de sistemas de energía, la tensión y la frecuencia se tratan de manera independiente. Sin embargo, en los sistemas de energía modernos, la integración de dispositivos basados en convertidores ha aumentado la interacción entre estas dos magnitudes. Por esta razón, ha crecido la necesidad de herramientas analíticas que cuantifiquen el impacto de los convertidores de energía en tanto la tensión como la frecuencia de dichas redes. En esta tesis, se propone una taxonomía de esquemas de control de convertidores de energía basada en el concepto de frecuencia compleja. Este enfoque captura las variaciones transitorias tanto de la tensión como de la frecuencia en una sola cantidad compleja. El marco propuesto permite desvincular el efecto en la tensión y la frecuencia de diferentes controladores participantes en estructuras de control complejas y la identificación de parámetros críticos, facilitando el diseño del control.

En esta tesis se presentan diferentes contribuciones relacionadas con la estabilidad y el control de tensión en sistemas de distribución activa. Para cada una de ellas, se presenta una revisión exhaustiva del estado del arte. El estudio investiga minuciosamente el impacto de la generación distribuida conectada a convertidores de energía en la red en la estabilidad de tensión de los sistemas de distribución modernos, al mismo tiempo que propone soluciones de control novedosas. Además, la investigación avanza más allá de los desarrollos teóricos al validar experimentalmente los análisis de estabilidad propuestos y al implementar y validar experimentalmente los esquemas de control en el Laboratorio de Integración de Energía Inteligente (SEIL) en el Instituto IMDEA Energía. Finalmente, la tesis concluye resumiendo sus hallazgos y ofreciendo pautas para futuras direcciones de investigación.

\selectlanguage{english}

	\tableofcontents
%\addcontentsline{toc}{chapter}{List of Figures}
%\listoffigures
%\addcontentsline{toc}{chapter}{List of Tables}
%\listoftables
\clearpage
\thispagestyle{headings}
%\addcontentsline{toc}{chapter}{List of Symbols}
%\printnomenclature[1.3in]
\newpage

\pagestyle{headings}

\mainmatter
	% Cabeceras 
	\pagestyle{fancy}
	\fancyhf{}
	\fancyhead[LE,RO]{\thepage}
	\fancyhead[RE,LO]{Chapter \thechapter. Introduction}
	
\chapter{Introduction}
\section{General Background and Motivation}
\subsection{Distributed Generation and Future Power Systems}
\label{subsec.CentralisedDecentralised}

\begin{figure}[t]
\centering
\includegraphics[width=0.9\columnwidth]{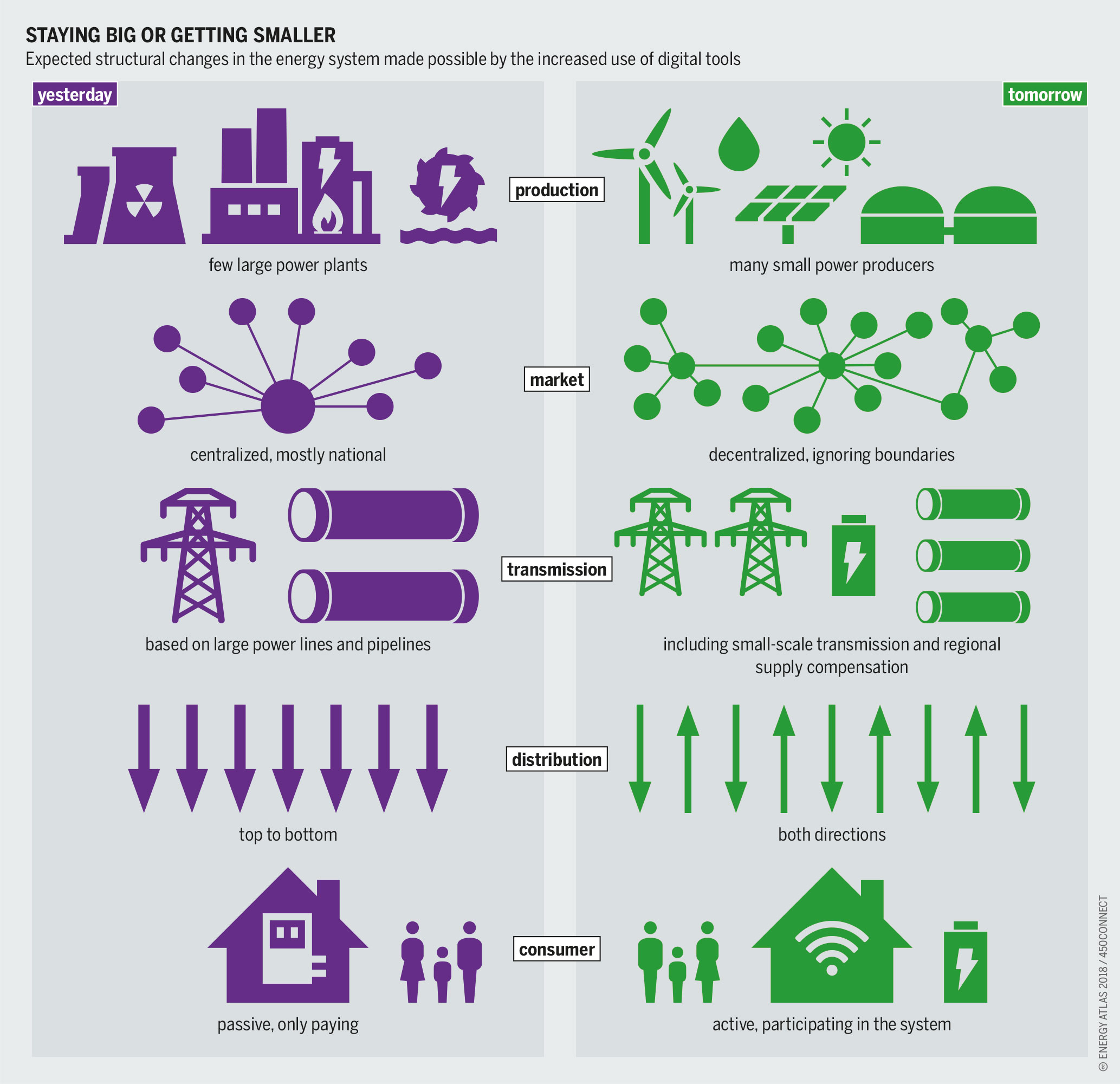}
\vspace{-0.1cm}
\caption{Comparison between the existing and future power networks. Reference: Graphic: Bartz/Stockmar, CC BY 4.0.}
\label{fig.future_grid}
\end{figure}

Power systems and electricity networks are currently in a process of rapid transformation.
From their conception and until a few decades ago, power systems had been operating solely in a centralised way.
Energy had been, almost exclusively, produced in large power plants operated by fossil fuels and then transmitted and distributed across the network.
However, with the introduction of renewable energy technology and the need for a rapid decarbonisation, a shift towards a more decentralised energy production, commonly referred to as \ac{dg} has occurred.
In this new paradigm, electrical energy is produced closer to the consumption and by generation devices whose number can scale up to millions~\cite{chatzivasileiadis2023micro}.
Figure~\ref{fig.future_grid} shows a schematic that compares the existing versus the future structure of power systems.

Several international directives and agreements have paved the way towards more sustainable energy production, with the EU Renewable Energy Directive and the Paris Agreement being notable examples~\cite{seo2017beyond}.
These efforts set out to mitigate the well-known problem of global warming and climate change, caused by fossil fuels and greenhouse gas emissions.
To this end, specific goals have been set regarding the decrease of fossil fuel emissions and the increase of the renewable energy share in the total electricity consumption~\cite{Roadmap_UE_2012,EU2020,EU2030}.
In order to comply with the international directives, several countries are set to increase their renewable energy production, with wind power and photovoltaic technology being on the forefront~\cite{renewables1,renewables2}.

During this profound change of the power networks, new technologies are increasingly taking a central role in their operation.
The intermittent and variable nature of renewable \ac{dg}, like wind and solar plants, has increased the necessity of energy storage to the networks~\cite{rahman2020assessment}.
Among the different storage technologies, batteries have been singled out as a reliable and mature solution~\cite{hesse2017lithium}.
At the same time, the electrification of the transport industry through electric vehicles is also introducing new challenges for the system operators~\cite{lopes2010integration}.
All of these technologies are interfaced to the grid via power electronic converters, highlighting the importance of their efficient and safe operation.

\ac{dg} introduces significant challenges to the network operation but also offers novel opportunities.
On the one hand, the stochastic nature of renewable energy sources increases system uncertainty~\cite{notton2018intermittent}.
In addition, \ac{dg} changes the way that electrical power flows through the network, compared to the conventional centralised generation.
This causes problems as legacy monitoring and protection equipment was not originally designed to operate in this way.
Finally, power electronic converters introduce new phenomena and disturbances to the system that may deteriorate the network operation and efficiency.
On the other hand, power converters offer fast and flexible control capabilities that can be leveraged to improve the reliability of the system~\cite{Iravani2010}.
Furthermore, by being located closer to the consumption, \ac{dg} can reduce transmission losses and overall improve system efficiency.
This combination of issues and opportunities that converter interfaced devices introduce to the system has attracted significant research attention and is the main motivation behind this thesis.

\subsection{Power System Stability - Traditional and Contemporary Approaches}

\begin{figure}[t]
\centering
\includegraphics[width=1\columnwidth]{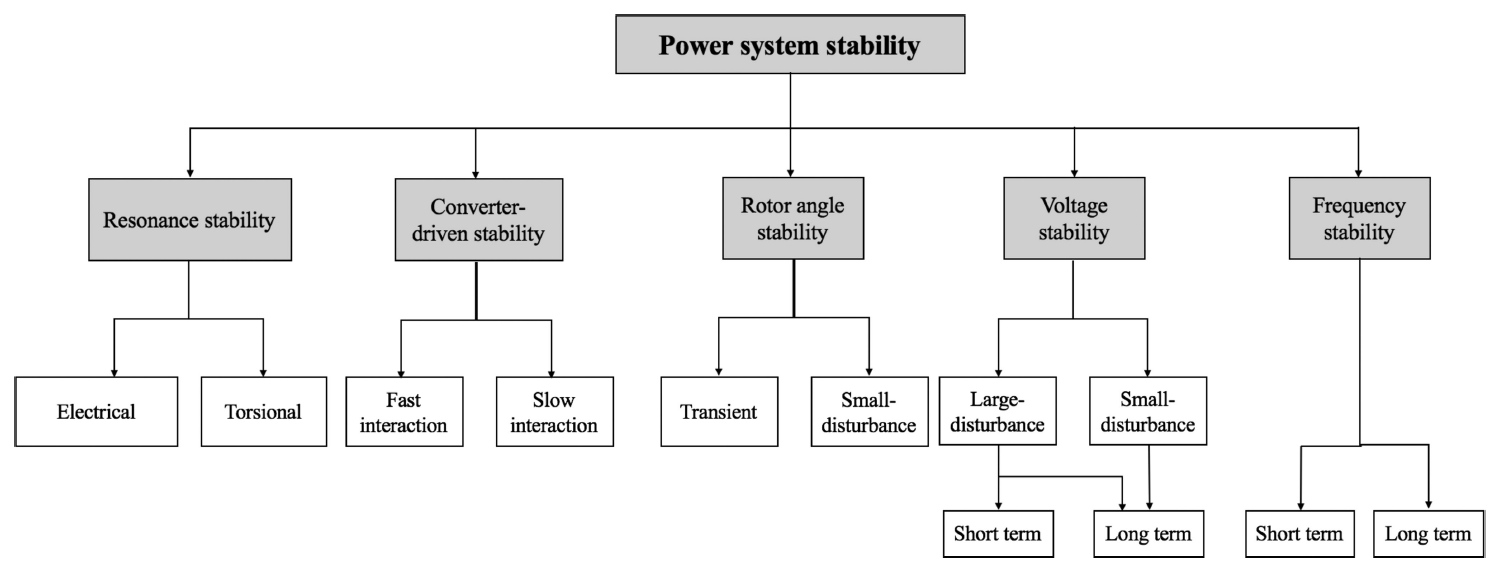}
\vspace{-0.1cm}
\caption{Power system stability classification~\cite{hatziargyriou2020definition}.}
\label{fig.stability_class}
\end{figure}

Power system stability has been a topic of study for several decades~\cite{kundur1994power}.
In~\cite{kundur2004definition}, a joint IEEE/CIGRE task force offered a definition of it, inspired by system theory, that can be readily applied and understood by a power systems audience.
In this work, power system stability was defined as \emph{the ability of an electric power system, for a given initial operating condition, to regain a state of operating equilibrium after being subjected to a physical disturbance, with most system variables bounded so that practically the entire system remains intact}.
The same work provided a systematic classification of different instability phenomena that occur in classical power systems.
This classification constitutes the traditional approach towards power system stability and contains the following stability categories~\cite{kundur2004definition}:

\begin{itemize}
    \item \textit{Rotor angle stability} refers to the capacity of synchronous machines within an interconnected power system to retain synchronism despite encountering disturbances. It relies on the ability of the system to establish or restore equilibrium between the electromagnetic torque and mechanical torque of each synchronous machine. In the case of instability, it manifests as an amplification of angular swings in certain generators, eventually resulting in their desynchronization from the rest of the system generators.

    \item \textit{Frequency stability} refers to the capability of a power system to uphold a consistent frequency subsequent to severe system disruptions which may cause a significant imbalance between generation and load. It depends on the ability of the system to regain or maintain equilibrium between the overall generation and load while minimizing unintentional load shedding. Instability for this case manifests as prolonged oscillations in frequency, leading to the tripping of generating units and/or loads.

    \item \textit{Voltage stability} refers to the ability of a power system to maintain steady voltage levels across all buses after being subjected to disturbances from the initial condition. It depends on maintaining or restoring equilibrium between the demand and supply of electrical power within the system. Instability arising in this case is characterized by progressive voltage drops or rises at certain buses. Such instability may result in the loss of load within a particular area or the activation of protective systems causing the tripping of transmission lines and other components, ultimately leading to cascading outages. The latter is commonly referred to as \textit{voltage collapse}.

\end{itemize}

The above categories were further divided into different sub-categories based on two criteria~\cite{kundur2004definition}.
The first criterion considered the timescale within which the instability occurs and the resulting stability types were labeled as \textit{short term} and \textit{long term} stability.
The second criterion considered the magnitude of the disturbance to which the system is subjected to.
For this criterion, stability can be divided to \textit{small-disturbance} stability and \textit{large-disturbance} stability.
The above definitions and classifications are valuable, not only because they shed light on the intrinsic mechanisms that cause instability but also because they provide insight regarding the suitable tools~(e.g. mathematical, computational, etc.) for the analysis of each phenomenon.

Since the proposal of the traditional power system stability classification, power systems have undergone significant changes, detailed in Section~\ref{subsec.CentralisedDecentralised}.
The increased penetration of converter interfaced \ac{dg} has led to the revision of the power system stability classification.
In~\cite{hatziargyriou2020definition}, an update of the different stability categories was proposed to account for all the changes that converter based systems have introduced.
Notably, two new stability categories were proposed.
They are the following:

\begin{itemize}
    
    \item \textit{Resonance stability} refers to cases for which energy is exchanged periodically and in an oscillatory manner. In case of insufficient energy dissipation, the magnitudes of electrical variables~(voltages, currents etc.) increase continuously, leading to instability. This category covers both electromechanical oscillations and purely electrical ones and as a consequence, it affects both conventional generators and converter interfaced generators. It can further be sub-divided to \textit{electrical} and \textit{torsional} resonance stability. The electromechanical, sub-synchronous resonances were intentionally left out of the initial classification of~\cite{kundur2004definition} due to the limited timescale that was considered in that work. However, the introduction of converter interfaced \ac{dg} has led to an expansion of the timescale of dynamic
    phenomena that are crucial for power system stability.
    
    \item \textit{Converter-driven stability} refers to cases that converter interfaced \ac{dg} units and their controllers are the main factors that affect the system dynamics and potentially, the loss of stability. Typically, this category includes the interactions of converters with the electromechanical dynamics of the machines, the electromagnetic transients of the network as well as the interaction between converters themselves. As a result, this category covers a wide range of phenomena and timescales. Based on their timescales, the interactions are labeled as \textit{fast} or \textit{slow} interactions.
\end{itemize}
Figure~\ref{fig.stability_class} shows a schematic of the updated and extended power system stability classification presented in~\cite{hatziargyriou2020definition}.
It should be noted that the above cases do not consider instabilities at the device level, but only network phenomena. 
In addition, stability of smaller and/or isolated networks~(e.g., microgrids) represents a separate topic~\cite{farrokhabadi2019microgrid}, that is out of the scope of this thesis.

The high complexity and the multitude of factors that affect converter-driven stability has warranted significant research attention in recent years.
However, the ways that converter based \ac{dg} affects voltage stability is unclear.
In~\cite{hatziargyriou2020definition}, only the impact of \ac{hvdc} to the short-term voltage stability was briefly considered.
However, it was noted that voltage stability and converter-based stability may overlap for the cases that power transfer between the converters and the rest of the system reaches its maximum limit.
This precise overlap, as well as the general contribution of power converters and \ac{dg} to voltage stability forms one of the main motivations of the present thesis.

\subsection{Voltage related Issues in Active Distribution Networks}
\begin{figure}[t]
\centering
\includegraphics[width=1\columnwidth]{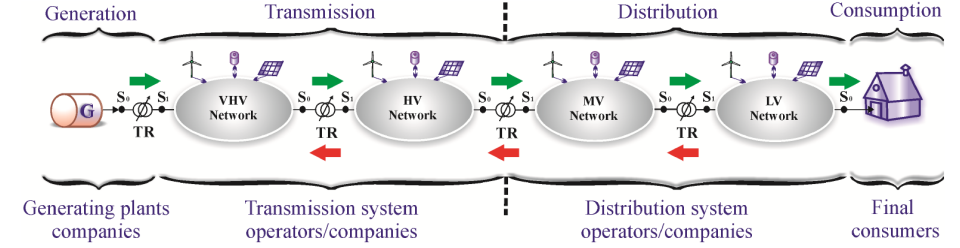}
\vspace{-0.1cm}
\caption{Schematic representation of reverse power flows from low voltage to higher voltage systems~\cite{gabash2016variable}.}
\label{fig.reverse_flow}
\end{figure}

\acp{adn} represent a paradigm shift in the way electrical energy is managed and distributed.
Conventional, passive distribution networks consist of relatively simple components like transformers and circuit breakers to deliver electricity with minimal intervention from the operators, producers and consumers.
During their operation, power predominantly flows in one direction, from the central generating station down to end-users. 
In contrast, \acp{adn} are characterized by their ability to dynamically manage power flows, integrating advanced technologies such as smart meters, sensors, and \ac{dg}.
These technologies provide adaptability and flexibility to the \acp{adn}, enabling greater resilience, efficiency, and responsiveness to the evolving energy landscape.
Furthermore, these networks empower consumers to play an active role in energy management, facilitate two-way power flows, and optimize grid operations in real-time~\cite{grvzanic2022prosumers}.
At the same time, these new technologies introduce several challenges to the network operation, with voltage stability related issues being the most prominent among them.

Voltage issues within \acp{adn} are a significant concern as these networks transition towards a more dynamic and decentralized structure.
The integration of renewable energy sources, such as solar panels and wind turbines, introduces an inherent variability into the system.
These sources generate electricity variably, causing voltage fluctuations that must be managed effectively to ensure a stable power supply~\cite{barker2000determining}.
In addition, when energy production from \ac{dg} exceeds the local consumption, the power flow is reversed which, in turn, leads to voltage rising above nominal operation levels~\cite{masters2002voltage}.
This presents a serious issue as voltage deviations beyond acceptable limits can lead to equipment damage, power quality issues, and even service interruptions, posing a direct threat to the reliability of the grid~\cite{olivier2015active}.
Figure~\ref{fig.reverse_flow} shows a schematic representation of how the flow of power is reversed in modern power systems under the presence of \acp{adn}.

Voltage related issues in \acp{adn} are aggravated by the fact that the conventional centralized voltage control mechanisms deployed in the existing grids become less adequate for the task.
These control mechanisms were designed and implemented for passive networks with few active elements and only downstream power flow from generation to consumption.
However, the ability of \ac{dg} to inject power into the grid during times of excess generation and absorb power during high demand, makes the design of smart controllers for their operation a relevant challenge~\cite{vovos2007centralized}.
The coordination of a number of \ac{dg} units to maintain voltage levels within specified limits is also a complex task, especially when information exchange and communication infrastructure are limited or unavailable.
The above topics, i.e. the mitigation of voltage fluctuations and rises in \acp{adn}, the design of advanced voltage controllers for \ac{dg} units and their coordination, all form an important motivation for this thesis.

% Voltage imbalance is another critical issue that arises in \acp{adn}~\cite{li2017optimal}.
% %
% Imbalances can result from unequal distribution of loads, disparities in \ac{dg} integration, or equipment malfunctions.
% %
% Voltage imbalances need to be carefully managed to ensure that all parts of the network receive power within acceptable voltage limits.

% In conclusion, voltage problems in Active Distribution Networks pose substantial challenges as the energy landscape evolves towards greater decentralization and renewable energy integration. Addressing these issues requires advanced control strategies, intelligent technologies, and collaboration among stakeholders to ensure the stability and reliability of the grid while maximizing the benefits of a more sustainable and resilient energy system.

\subsection{Power Converter Control for Distributed Generation}

\begin{figure}[t]
\centering
\includegraphics[width=1\columnwidth]{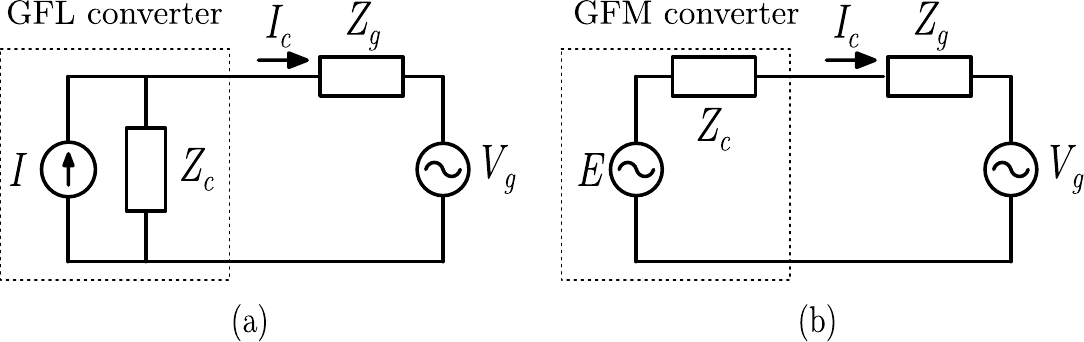}
\vspace{-0.1cm}
\caption{Suitable  representations of (a) a GFL converter and (b) a GFM converter, using their electrical circuit equivalents~\cite{rocabert2012control,rosso2021grid}.}
\label{fig.gfm_gfl}
\end{figure}

The large increase of \ac{dg} in modern power networks has led to a variety of strategies for their control~\cite{blaabjerg2006overview}.
The high controllability and fast response times of power converters offer a useful tool that can be leveraged for the improvement of power grid operation and the increase of renewable energy generation.
Currently, the prevalent approach is to group different converter control schemes into the \ac{gfl} or \ac{gfm} categories~\cite{rocabert2012control, li2022revisiting}.
The former are meant to measure or estimate the angle of the grid while passively injecting the power that is requested from them.
The synchronization is usually achieved by using a \ac{pll} control block~\cite{li2022revisiting, wang2020grid}.
Extensive research regarding \ac{pll} stability and synchronization capability, especially in weak grids, has been carried out~\cite{wu2022synchronization, chen2022impact}.
%
%However, it is commonly assumed that \ac{gfl} converters with \acp{pll} do not affect the frequency at their connection point.
%
\ac{gfm} converters do not rely on the estimation of the grid voltage angle. 
Instead, they achieve synchronization with the grid by varying their active power injection~\cite{wang2020grid}.
A fundamental controller for both configurations is the \ac{cc}.
It serves to accurately control the output current of the converter and to limit its value in the case of contingencies, a critical feature for the safe operation of the converter~\cite{timbus2009evaluation}.
A critical difference between \ac{gfm} and \ac{gfl} control configurations is whether a PI-based voltage controller is cascaded with the \ac{cc}.
The inclusion of this voltage controller or not dictates the operation of the converter and its stability, especially when connected to grids of different strength~\cite{chen2016impedance}.
It also decides on whether the converter operation can be better approximated by a current source or by a voltage source.
Figure~\ref{fig.gfm_gfl} shows two simplified representations of \ac{gfl} and \ac{gfm} converters.
They are depicted as a current source connected in parallel with an admittance and as a voltage source connected in series with an impedance, respectively~\cite{rosso2021grid}.

Various power-based synchronization strategies can be found in the literature~\cite{wang2020grid, tayyebi2020frequency, rosso2021grid}.
Despite the choice of \ac{gfm} control method formulation still being an open research question, the most common ones found in the literature and the practical applications are \emph{droop control} and \emph{\ac{vsm}}~\cite{4118327, moran2021influence, zhong2010synchronverters, roldan2019design, gonzalez2021design}.
Droop controlled converters adjust their active power outputs using an active power-frequency ($P/f$) function to achieve synchronization~\cite{4118327, moran2021influence}.
\ac{vsm}-based converters fully emulate a swing equation within their control structure, typically of second order~\cite{zhong2010synchronverters, roldan2019design, gonzalez2021design}.
Many \ac{vsm} implementations exist with all recent works highlighting the necessity of inner current and voltage controllers in a cascaded configuration~\cite{d2013virtual}.
%
%These cascaded loops allow the explicit inclusion of voltage and current limitations that are necessary for the safe operation of the converters~\cite{d2015virtual}.
%
%Moreover, dedicated voltage control loops are required for controlling converters as voltage sources~\cite{wang2020grid}.
%

Quantifying the effect of all the internal~(current, voltage, \ac{pll}) and external~(droops, \ac{vsm}) controllers on the voltage and the frequency at the converter connection point is of interest to both research and applied engineering communities.
This can lead to a better understanding of the link between the controller design and the network stability.
In addition, the adaptation of existing controllers and the design of novel controllers for maximizing renewable energy integration to the networks present a relevant challenge for the future of the energy transition.

\subsection{Main Objective and Scope of the Thesis}

The increased presence of converter based \ac{dg} to power systems introduces at the same time  technical challenges and opportunities.
More specifically, the problem of preserving stability in power systems with high penetration of power electronic converters is of particular importance for their safe and reliable operation.
Out of different power system stability categories, the research in voltage stability is the most relevant for weak and resistive distribution networks under the presence of power converters.
Also, the mitigation of voltage deviation problems in active distribution networks is critical, since operation outside of the specified voltage limits can lead to equipment disconnection and potentially, equipment damage.

Power converters and their control system are an effective tool to address voltage related issues in modern power networks.
The flexibility and fast response time of converters can be used to reduce voltage deviations and increase stability margins.
The main objective of this thesis is to assess the effect of power converter based \ac{dg} on the stability of \acp{adn} and to propose novel controllers for the mitigation of voltage-related issues in such networks.
A secondary objective is to investigate the effect of different power converter controllers on transient voltage phenomena.
%

%To address these principal objectives, the main tasks performed and detailed in this thesis are the following:

The principal objectives of the thesis can be divided further into the following 
ancillary objectives:

\begin{itemize}
    \item Review the state of the art for stability analysis techniques of power systems with high penetration of power electronic converters.
    \item Model power system dynamics and converter control dynamics for the purpose of voltage stability analysis and assessment.
    \item Apply bifurcation theory for the evaluation of voltage stability and parameter stability margins for \acp{adn}, with emphasis on the interaction between power converters with other voltage regulating elements of the network
    %(e.g. \acp{ltc})
    and on the limitations that their finite capacity imposes on their ability to regulate voltage.
    \item Review the state of the art of voltage controllers for power electronic converters in electrical distribution systems.
    \item Develop a novel voltage controller  to reduce voltage deviations in \acp{adn} by taking into account the active and reactive power coupling in resistive networks. Consider its implementation across multiple control layers, both local and centralised.
    \item Assess and quantify analytically the impact of different control scheme topologies on transient voltage and frequency.

    \end{itemize}

% The scenario described in the previous section depicts a power system where RES and other elements are connected to the grid by using power electronics converters representing a sizeable share of generation units. \textbf{The objective of this thesis is to investigate and evaluate the impact of a massive deployment of power electronics interfaces on the stability of power systems and to propose control system improvements for the grid-connected converters that safeguard the stability of future power networks.} The specific tasks to address this objective include:
% %
% \begin{itemize}

% \item Detailed modelling of the specific dynamics introduced by power electronics converters and stability analysis under weak grid conditions.

% \item Detailed modelling of control techniques based on the emulation of synchronous generators and analysing their stability under weak grid conditions.

% \item Improve control algorithms for conventional converter topologies used in the integration of RES. This includes high performance control systems featuring robustness under weak grid conditions, voltage disturbances, etc.

% \item Improve the control algorithms for HVDC transmission systems used to interconnect low inertia networks. This includes control algorithms that provide voltage and frequency support and power oscillation damping.

% \item Evaluate the stability of MTDC networks and propose control algorithms for VSCs operating in hybrid AC and DC networks.

% \end{itemize}

\section{State-Of-The-Art}

\subsection{Power system stability analysis methods}

There are several methods for voltage stability evaluation in power systems~\cite{van2007voltage}.
The most prominent of them are categorized to small-, large-signal and continuation methods.
All of them present unique capabilities and drawbacks and will be discussed in the following section.

Small-signal techniques assume linearization of the original nonlinear model, representing the complete power system, around an operating point.
This allows the use of mathematical tools from linear analysis for the evaluation of the system stability.
The two most relevant small-signal methods are the eigenvalue analysis~\cite{milano2020eigenvalue} and the impedance criterion~\cite{sun2009small}.
The eigenvalue analysis is based on the calculation of the eigenvalues of the matrix representing the linearized model and then, depending on their values, determining system stability.
Apart from being able to decide on whether the system is stable or not, the eigenvalue method provides additional information regarding the damping and oscillatory frequency of critical eigenvalues~\cite{Wang2016}.
Also, through participation factor analysis, it provides quantitative links between the critical eigenvalues and their related states and algebraic variables~\cite{pagola1989sensitivities,tzounas2019modal}.
Despite its significant merits and its widespread use, the eigenvalue method requires the full model of the studied system, together with the calculation of its Jacobian matrix, two elements that are not always readily available.
More importantly, due to the linearisation step that is essential for it, the eigenvalue method can only provide stability information for the operation point, or for regions relatively close to it.
For this reason, it is not suitable for studies in which the system variables or parameters vary significantly from that point.
Another common method based on linearisation is the impedance method~\cite{sun2011impedance}.
It consists of splitting the power system into two distinct sub-systems at an arbitrary point.
Then, the current-to-voltage transfer function, also known as the impedance transfer function or simply the impedance, for each sub-system is derived.
Finally, the Nyquist criterion is applied to the ratio of the two impedances to decide on the stability of the system~\cite{desoer1980generalized}.
The main advantage of the impedance method is that the necessary impedance transfer functions, apart from being derived analytically, can also be accurately estimated from measurements~\cite{wang2019small}.
This allows the incorporation of new devices to networks that are already in operation, whose models may not be readily available, without jeopardizing their stability.
Furthermore, the impedance method provides information with respect to the operation of the network in a wide frequency range while also providing the system stability margins~\cite{mateu2023oscillatory}.
The above merits have led to the widespread use of the impedance method in different applications, ranging from dc-dc converters to wind-farms and HVDC~\cite{amin2019nyquist}.
However, being a linearisation-based method, in a similar way as the eigenvalue method, it fails to predict the system stability when parameters and variables diverge significantly from their nominal values.

Large-signal stability analysis methods do not rely on linearisation and instead use nonlinear mathematical models for the system.
The use of nonlinear models makes these methods more suitable for the study of large transients in power systems, like for example the case of sudden voltage dips.
Large-signal methods are based on advanced mathematical tools like dynamical systems theory and Lyapunov theory~\cite{khalil02}.
Specifically, a common approach is to define a function that represents the total energy of the system and by studying its properties, e.g. studying the values for which the function and its derivatives take positive/negative values, define the stability boundaries of the system.
These functions are known as energy-based or Lyapunov functions.
When these functions meet specific criteria, global stability for the system can be guaranteed.
Lyapunov or energy-based methods have been extensively used in the past for the study of classical, synchronous machine-based systems~\cite{pai1989energy,fouad1991power}.
Recently, their application in stability studies of microgrids has also received significant attention~\cite{Kabalan2017}.
From the perspective of \ac{vsc} control, both \ac{gfm} and \ac{gfl} control structures have been studied and compared by using these methods~\cite{fu2020large}.
The synchronization of the \ac{pll} has also been studied by using Lyapunov-based methods~\cite{zhang2021large}.
Despite the merits of the approach, its applicability to large networks with different types of converter interfaced devices is limited.
The reason behind that is that finding suitable Lyapunov functions is a very complex procedure and their existence for certain cases is not guaranteed~\cite{chang1995direct}.
In most cases, one has to resort to Lyapunov candidate functions that are system- or control-specific.
An algorithmic approach to formally construct Lyapunov functions for general power system topologies and different connected devices, despite significant research efforts in the past, is an ongoing research topic~\cite{anghel2013algorithmic}.

Continuation-based or homotopy methods are methods to calculate the stability limits of dynamic systems by varying continuously one or more parameters of such a system~\cite{seydel1988equilibrium}.
Compared to simply changing the parameter of a linearised system, in the continuation method the selected parameter to be varied is incorporated to the system as a variable~\cite{milano2010power}.
This increases the robustness of the method and guarantees the convergence of the algorithm, even for large deviations from the original value of the parameter~\cite{ortega2000iterative}.
For this method, the selected parameter is assumed to be changing in a smooth, continuous way~\cite{arnol2003catastrophe}.
%
%These variations are assumed to occur ``slowly'', while the system equilibrium is preserved.
%
This assumption is also known as the quasi-static assumption~\cite{avalos2008equivalency}
In power systems, a specific variation of the continuation method, called continuation power flow, has been used extensively for the calculation of the maximum loading point of such systems~\cite{ajjarapu1992continuation,hiskens1995analysis}.
For this specific case, the selected parameter is a scalar quantity called the loading factor, multiplying all the load values and the active power generation of the system~\cite{milano2010power}.
The maximum loading point of a system has been identified as closely associated with the point of voltage collapse, rendering it an important characteristic of the system for voltage stability studies~\cite{van2007voltage}.
Since its introduction, the use of continuation power flow has been extended for hybrid, ac/dc systems~\cite{canizares1993point} and radial distribution systems~\cite{dukpa2009application}.
Recently, an adapted continuation power flow algorithm that considers discontinuous events was proposed~\cite{colombari2019continuation}.
Although the use of continuation methods have been used extensively in traditional power systems, their application to power systems with a high penetration of power converters has been, thus far, limited.
For this reason, the adaptation of continuation methods for the stability analysis of converter-dominated electrical power systems is an ongoing research topic.
\subsection{Bifurcation Analysis of Power Systems}

Bifurcation theory is the mathematical study of changes in the qualitative or topological structure of a given family of differential equations~\cite{hirsch2012differential}.
It provides the theoretical background to understand and classify the rich variety of behaviours that can be encountered in nonlinear dynamical systems~\cite{strogatz2018nonlinear}.
Specifically, bifurcation theory deals with one key aspect of nonlinear systems, namely the emergence of sudden changes in system response arising from smooth, continuous parameter variations~\cite{arnol2003catastrophe}.
These variations can be multi-variable in the general case, but for the sake of simplicity single parameter variation techniques are often used~\cite{seydel2009practical}.
The pairing of bifurcation theory with continuation techniques provide a strong tool, with both qualitative and quantitative characteristics, for the stability assessment of power systems.
The combination between the computation of stability limits with continuation techniques and their classification based on bifurcation theory is also known as \emph{bifurcation analysis}.

Considering the inherent non-linearity of the power flow equations, bifurcation theory was firstly proposed in the 80's for the study of voltage collapse in power systems~\cite{kopell1982chaotic}.
Among the rich variety of bifurcations in dynamic systems, the ones encountered mostly in power systems are \acp{snb} and \acp{hb}~\cite{van2007voltage}.
Saddle node bifurcations appear when a couple of equilibrium points, one stable and one unstable, merge at the bifurcation point and then locally disappear.
These bifurcations are also associated with a real eigenvalue of the system state matrix crossing the imaginary axis.
In classical power systems, \acp{snb} are related with the voltage collapse point of the system, which also often coincides with the point of maximum system loading~\cite{canizares2002voltage}.
\acp{hb} appear when an equilibrium solution becomes a periodic solution.
These periodic solutions of the dynamic system are also known as limit cycles and can be either stable or unstable.
\acp{hb} are also associated with the cases for which a pair of complex conjugate eigenvalues of the state matrix crosses the imaginary axis, resulting in the emergence of oscillations in the system.
Other bifurcations that, less commonly, appear are \acp{lib} and \acp{sib}~\cite{ma2018voltage}.
The former are closely associated in traditional power systems when conventional generators reach their reactive power limits~\cite{milano2010power}.

\subsubsection{Bifurcation Analysis for Traditional Power Systems}

In traditional power systems, bifurcation analysis has been applied for the calculation and classification of stability margins, mainly focused on the voltage collapse induced through \acp{snb}~\cite{dobson1989towards}.
In these studies, emphasis was given in defining the conditions for a \ac{snb} to occur~\cite{canizares1995conditions}.
In~\cite{canizares1995bifurcations}, the effect of load modelling on the \ac{snb} occurrence was investigated.
In \cite{ajjarapu1992bifurcation}, the importance of nonlinear analysis was emphasized, especially on heavily stressed networks, while in~\cite{revel2009bifurcation}, bifurcation analysis was conducted in a multi-machine system.
More recently, the existence of both local and global bifurcations was reported in a small power system~\cite{sakellaridis2011local}.
%
%Bifurcation analysis has also been used for the design of nonlinear controllers~\cite{lee2001study,vahdati2016hopf,vahdati2017hopf2}

\subsubsection{Bifurcation Analysis of Converter-Based Devices}

The introduction of various, converter-based devices to modern power networks rejuvenated the use of bifurcation analysis for the study of the non-linear effects dictating their operation.
In the work of \emph{Huang et al}~\cite{huang2013catastrophic,huang2017bifurcation,huang2013low,wan2015effects}, analytical formulations of bifurcation phenomena in grid-connected \acp{vsc} is carried out.
Catastrophic bifurcations, i.e. bifurcations after which normal operation cannot be re-established, are identified for both the DC- and AC-side of a converter, following large variations in the operating conditions ~\cite{huang2013catastrophic,huang2017bifurcation}.
The physical origin behind both types of instability is revealed, using full and reduced-order models.
Boundaries for these catastrophic bifurcations are derived in the parameter space.
In~\cite{wan2015effects}, the shrinking of the stability area due to the existence of multiple converters and their interaction through a non-ideal grid is studied.
In~\cite{huang2013low} the non-ideality of the grid is reported to cause a low-frequency \ac{hb} in a three-phase power-factor-correction power supply.
The effect of the current limiter on the stability margin and the potential loss of stability through a \ac{lib} were discussed in~\cite{hu2017bifurcation,xing2021limit}.
Lastly in~\cite{yang2021homoclinic,yang2023comparison}, homoclinic bifurcations for \ac{gfl} and \ac{gfm} converters were reported and compared.
These studies highlight that the operation of converter interfaced devices affect the stability of power systems in unique ways, that were not previously considered.
This can be of concern for the safe operation of the network, particularly since, in many cases, instability occurs while the system is close to seemingly safe operating conditions.

Bifurcation analysis has been applied to various others converter-based components, different than the regular converter interfaced \ac{dg}.
In~\cite{yang2022bifurcations}, bifurcation-induced instabilities in \ac{gfl} rectifiers were studied.
Bifurcation-related instabilities in doubly-fed induction generator wind farms, caused from control and network parameter variations were studied in~\cite{yang2011oscillatory,khosravi2018analysis}.
The dynamic effect on the system stability of wind farms based on permanent magnet synchronous generator configuration was studied in~\cite{6742735}.
\ac{hb} and Neimark bifurcations were reported after the analysis of \ac{vsc}-based, grid-connected STATCOMs~\cite{skandarama2015control,5109838}.
These studies confirm that bifurcation phenomena appear on various converter-related devices, jeopardizing stability.
 
\subsubsection{Bifurcation Analysis of Systems with high Converter Participation and Microgrids}

On the system level, significant efforts have been dedicated to microgrid stability analyses using bifurcation theory.
In~\cite{lenz2018bifurcation}, a microgrid consisting of two inductive droop inverters and several time-invariant nonlinear dynamical constant power loads was studied.
The main dynamic phenomena that occur when the power output varies were analyzed, stability margins for different parameters were evaluated and design oriented diagrams were provided.
Similar analyses can be found in~\cite{shuai2019parameter}, considering \ac{im} and \ac{zip} loads.
Finally, a dynamic stability analysis of microgrids dominated by converters emulating synchronous machines can be found in~\cite{shuai2017dynamic}.

In~\cite{sreeram2017hopf} the case of sudden instability occurring in supposedly stable regions of the microgrid operation was highlighted.
%
%On the contrary, \ac{hb} analysis on various-sized microgrids showed that the region marked as stable in parameter space is stable irrespective of the size of perturbation.
%
In the work of \emph{Diaz et al}~\cite{diaz2010composite,diaz2010scheduling} bifurcations in isolated microgrids are discussed.
Specifically, bifurcation analysis is used in \cite{diaz2010composite} for determining the loadability margin of the microgrid.
In~\cite{diaz2010scheduling} a bifurcation analysis-based procedure is proposed to determine droop coefficient selection and primary reserve scheduling.
The security region of these droop parameters is discussed in~\cite{wu2014small}.
Lastly, in~\cite{chiang2015available} local bifurcations are treated as system constraints, among node voltage limits and thermal limits, in order to calculate the available delivery capability of a distribution network.

Bifurcations have been used for the stability analysis of DC microgrids~\cite{benadero2015nonlinear}.
The nonlinear analysis of a droop controlled DC microgrid in provided analytical stability conditions and a guideline for droop gain selection~\cite{tahim2015modeling}.
Finally, bifurcations were identified and discussed in photovoltaic-battery hybrid power system~\cite{huang2017bifurcationhybrid,xiong2013bifurcation,xiong2015bifurcation}. 
In such systems, non-smooth bifurcations often occur~\cite{xiong2019non}.
Despite the fact there have been many  applications of bifurcation analysis in converters and islanded microgrids, grid-connected microgrids and electrical distribution systems with high penetration of power electronic converters have received less attention.

\subsection{Circuit-Inspired Control Loops}

A current control loop is fundamental for the operation of most grid-connected DG units. 
However, its negative impact on the grid stability has fostered the incorporation of circuit-inspired control loops in the control systems of DG units.
%has been a topic of extensive study. 
%
Specifically, the virtual impedance has been widely used to increase robustness and power-sharing between parallel DG units, mainly in microgrids, where DGs are operated as voltage sources~\cite{he2011analysis}. 
Different implementations have been proposed in the literature~\cite{wang2014virtual}. 
An enhanced virtual impedance controller was proposed in~\cite{he2013islanding} in order to achieve accurate reactive and harmonic power sharing.
In~\cite{mahmood2014accurate}, an adaptive implementation for the virtual impedance was proposed to account for the changes in the loading of the system.
Recently, instead of designing controllers based on well-known circuit concepts, attempts have been made to interpret and visualize standard controllers as control circuits.
In~\cite{li2020impedance}, the inner- and outer-controllers of a \ac{gfm} converter were visualised and analyzed as circuit elements while in~\cite{fan2022equivalent}, an equivalent variable resistor model was derived to model a circular current limiter in a \ac{gfm} converter.

The dual concept of the virtual impedance for current-controlled converters is the virtual admittance. 
It was first introduced in~\cite{rodriguez2013control}, and it has been mainly used to emulate the equivalent inductance of synchronous machines~\cite{rodriguez2013control,rodriguez2018flexible,zhang2016frequency}
and for harmonic compensation~\cite{blanco2016virtual,micallef2015mitigation,tarraso2017grid}. 
This technique does not require derivatives of any signal, thus making its implementation straightforward and robust against noise~\cite{rodriguez2013control}. 
Recently, the concept of virtual admittance has been applied to a novel \ac{gfm} controller for weak grids that abolishes the need for the use of PI-based voltage controllers~\cite{leon2023grid}.
However, the potential contribution of the virtual admittance to the voltage regulation in \ac{gfl}, \ac{dg} units has not been sufficiently explored.
% to the extend of the authors knowledge
%

The design and implementation of circuit-inspired controllers need to be carefully considered to account for voltage stability, resonances and parametric instabilities~\cite{braslavsky2017voltage,mohamed2010mitigation}.
Hence, robust implementations of such controllers that allow a wide range of parameter selection are of interest.
In practical implementations of the virtual impedance controllers, the necessity to avoid derivation of high frequency noise resulted in the quasi-stationary, or algebraic, approximation~\cite{8693833}. 
Since then, this approximation has been successfully applied to the design of harmonic virtual admittances~\cite{9376914} and machines~\cite{mo2016evaluation}.
Results show that the approximation successfully reproduces the steady-state operation while at same time offers improved damping and performance during transient conditions.
The application of the quasi-stationary approximation to a virtual admittance based voltage controller has not been explored in the literature.

\subsection{Voltage Control for Distributed Generators}

In recent years, \ac{dg} has been established as an attractive way of providing electricity to power networks and has been increasingly used to replace conventional generation~\cite{bollen2011integration,guerrero2010distributed}. 
\ac{dg} allows the connection of renewable energy sources while promoting self-consumption and local power generation, thus paving the way towards the decarbonization of electrical grids.
However, \ac{dg} has been also linked with detrimental phenomena for the power grids, like the presence of both overvoltage and undervoltage conditions~\cite{macken2004mitigation,carvalho2008distributed}.
%
%It has also been reported that they might introduce resonance effects in the grids~\cite{wang2013resonance}.
%
In the past, \ac{dg} units were expected to either preserve their normal operation during grid voltage variations or to disconnect in the case of severe disturbances.
However, as \acp{dg} units are taking a larger share of demand coverage in the grid, the expectations and requests for them to actively contribute to the grid regulation has risen in recent years~\cite{yang2013suggested,8332112}.
Specifically, voltage support is considered one of the main requirements for \ac{dg} integration in electrical distribution systems~\cite{braun2012distribution}.
Several voltage regulating strategies have already been proposed for \ac{dg} units.
The classification of voltage regulation control schemes for \ac{dg} can be decided mainly by two different criteria.
The first one is on whether they adjust the active power injection of the converter, the reactive one or both.
The second criterion concerns the type of information available to each converter and the communication requirements of the control.
Based on this criterion, they can be categorized between local, centralised, decentralised controllers as well as hybrid combinations of the above~\cite{evangelopoulos2016optimal}.

\subsubsection{Volt/VAR Control}

The standard method to control the voltage in highly inductive networks is through reactive power, i.e. implementing a Volt/VAR control~\cite{8964320,9656605}.
Typical high-voltage, transmission networks fall into this category.
Such networks present a strong coupling between reactive power and voltage, making this type of controllers highly efficient~\cite{vasquez2009voltage}.
Several control schemes for reactive power management that provide voltage support during faults have been proposed for generic \ac{dg} units~\cite{camacho2012flexible,castilla2013voltage,calderaro2014optimal}.
Specific voltage control solutions utilising reactive power for different devices have also been proposed.
These devices include wind-farms, photovoltaics and even constant power loads~\cite{4349135,mastromauro2012control,jelani2013reactive}.
The stability of such controllers has been addressed in the past~\cite{andren2015stability,braslavsky2017voltage}, while the optimal tuning of Volt/VAR controllers for the whole network is a topic that is currently receiving attention~\cite{murzakhanov2023optimal,gupta2023deep}.
Finally, many publications have tackled the problem of the optimal placement of  reactive power sources~\cite{sun2019review}.
Several algorithms have been proposed, utilising concepts such as trajectory sensitivity index~\cite{5299293}, mixed integer
programming~\cite{4839926} or the minimization of the empirical controllability covariance~\cite{7450656}.
An algorithm for placement of DGs specifically for voltage reliability purposes is provided in~\cite{ettehadi2013voltage}.
%
%Other DG optimal placement strategies based on different criteria can be found in the literature as well~\cite{shaaban2012dg,hejazi2012independent,gautam2007optimal,acharya2006analytical}.
%
The above references provide a variety of examples on the use of Volt/VAR controllers and reactive power management for voltage support.
However, for resistive and mixed networks, the development of novel controllers that manage both active and reactive power towards the aim of voltage support is an open research topic.

\subsubsection{Volt/Watt Control}

In low-voltage networks, the high $R/X$ ratio makes reactive power control less effective for voltage regulation~\cite{kabiri2014influence}.
For this reason, voltage control through active power management is also considered.
In resistive microgrids, where converters are responsible for forming the grid, this is achieved with the so-called inverse droops~\cite{de2007voltage}.
This strategy consists in reversing the standard $P/f$ and $Q/V$ droops and in using complementary $Q/f$ and $P/V$ droops instead.
For active distribution networks with high penetration of \ac{dg} sources, controllers using Volt/Watt functions are considered~\cite{tonkoski2010coordinated,kashani2018smart}.
These functions typically call for active power curtailment of the \ac{dg} sources, most commonly photovoltaics, in the case of overvoltage violation.
Due to the additional requirement for active power reserves, active power injection during undervoltage scenarios is less common.
Due to the negative financial impact of active power curtailment for producers, different strategies to achieve fair active power curtailment among produces along a distribution network have been developed~\cite{liu2020fairness,gerdroodbari2021decentralized}.
Overall, the cost of active power curtailment and of the maintenance of active power reserves calls for an efficient management of converter capacities.
To this end, controllers that affect both the active and reactive power output of the converter are of interest.

\subsubsection{Combined Active and Reactive Power Voltage Control}

In recent years, because of the proliferation of converter interfaced \ac{dg}, the natural coupling that exists between the different variables of a power system~(e.g. active and reactive power injections, frequency and voltage) has been magnified.
For this reason, combined use of active and reactive power towards the goal of voltage regulation has been gaining ground~\cite{6755528,kawabe2017novel,zhong2022improving}.
In~\cite{6755528}, active and reactive power droop functions were used to regulate voltage locally, with the droop values being updated by a central controller.
In~\cite{kawabe2017novel}, a dynamic voltage support strategy was proposed that equally prioritised active and reactive power injections for voltage dip mitigation.
%
%In~\cite{zhong2022improving}, active and reactive power were used simultaneously to regulate both frequency and voltage during transients in a high-voltage, transmission network.
%

In distribution networks, in order to maximize the voltage regulation capability of \ac{dg} units, a combination of active and reactive power is utilised~\cite{collins2015real,fusco2021decentralized}.
In~\cite{olivier2015active}, reactive power injection is prioritised during overvoltages and active power curtailment by the \ac{dg} unit is required only when the reactive power limits have been depleted.
The performance, effectiveness and impact of these strategies on the grid stability have already been addressed in the literature~\cite{braslavsky2017voltage}.
In all the studies mentioned above, active and reactive powers have been controlled independently. 
This means that the coupling between the two in low-voltage networks has been scantly explored and it is generally considered to be a side effect~\cite{zhong2022improving}.
Recently,  in~\cite{zhong2022improving}, $P-Q$ coupling has been exploited in the control of distributed energy resources to improve the dynamic response of power systems.
However, transient conditions were prioritised over steady state voltage deviations and only simplified models of the current controllers and the power converters were used.
%

% %%% Motivation

% The motivation for applying the virtual admittance control for voltage profile improvement stems from the \ac{dg} ability to operate seamlessly between inductive and capacitive modes while also adjusting its active power output.
% %
% This resembles an operation of a variable admittance and extends the capabilities of capacitor banks, a standard control measure against undervoltage in traditional power systems~\cite{aman2014optimum}.
% %
% This operation can contribute to balancing generation/load consumption and thus, by mitigating line congestions, present a solution to the fundamental issues behind the control of voltage in power systems~\cite{ORBi-30326279-329a-427d-a9d3-c67606fa5602}.
% %
% The potential contribution of the virtual admittance controller (VAC) to the voltage regulation from \ac{dg} units was briefly explored in~\cite{moutevelis2021virtual}.
% %
% In this work, only the effect of one \ac{dg} was studied.
% %
% However, to study effects specific for each DG and its connection point~(e.g., distribution lines mismatch and different loading for each system node) multi-converter system configurations must be taken into consideration.
% %
% Also, it is necessary to assess the impact of the controller on the overall system, not just at the DG connection point.
% %

%%%%

\subsubsection{Local Control}

Local controllers for voltage regulation use measurements only from the point of connection of the \ac{dg}.
They are a reliable and easy to deploy solution as all the necessary sensors and measurements are offered by the converter, thus not requiring any communication infrastructure.
Historical data regarding the electrical characteristics of the connection point are often used to tune their control parameters and their settings are typically not altered after installation.
However, this data might be unavailable and the electrical characteristics might be changing over time, especially with the further installation of more devices~\cite{weckx2016optimal}.
Furthermore, it has been reported that corrective actions to improve the voltage profile locally can have an adverse effect in other nodes of the grid~\cite{fusco2021decentralized}.
For the above reasons, a complementary control layer that monitors the voltage level of the whole network and properly adjusts the control parameters is beneficial.

\subsubsection{Centralised Control}

Centralised controllers use measurements from the whole grid and provide control set-points to each distributed generator~\cite{kim2012coordinated}. 
The main objective is to adjust the power injection to the system nodes in order to match the load consumption and thus, regulate their voltage.
The motivation is similar to the peak reduction and valley filling algorithms developed for electric vehicles~\cite{liang2018dynamic}.
The advantage of centralised methods is that, by observing the full state of the network, they can calculate the optimal operation point for the full grid.
Their downsides are that they require extensive grid metering and a robust communication network while also being vulnerable to cyberattacks~\cite{fusco2021decentralized}. 
Typically for distribution networks, reliable metering and communication infrastructure is not available and for this reason, a robust control layer that exclusively depends on local measurements is favourable. 

\subsubsection{Decentralised Control}

Decentralised controllers can refer to a wide variety of controllers that require minimal communication~\cite{gerdroodbari2021decentralized}.
Hence they constitute a distinct category that is neither fully local, nor completely centralised.
Communication can be limited only to devices connected to the same phase, to devices connected to neighbouring nodes, or solely to low-bandwidth power line communication~\cite{olivier2015active}.
Their objectives may range from power loss reduction, overvoltage prevention or fairness in active power curtailment.
One typical example of decentralised control that considers communication capabilities between the devices but lacks a centralised control structure is the agent-based control approach~\cite{jennings2003agent}.
Agent-based systems comprise of autonomous entities, called agents, that act flexibly in order to achieve their objectives.
They can either operate autonomously or exchange limited information with each other.
Since their introduction, agent-based systems have been successfully applied for the voltage regulation in smart grids~\cite{aquino2011control,zhang2015agent}.
Lastly, hybrid approaches combining local and central control methods have been proposed~\cite{weckx2014combined}. 
Decentralised and hybrid methods take advantage of the robustness and reliability of local methods as well as make use of possible communication infrastructure to improve the performance of the whole network.
For this reason, they are considered a viable solution until fully monitored distribution grids with highly reliable, high bandwidth communication channels are available.

\subsection{Complex Frequency: A new concept for power system analysis and control}

The physical meaning and the precise definition of the instantaneous frequency of a signal has been the focus of several discussions over the years~\cite{van1946fundamental}.
Specifically, the definition of the frequency in a three-phase power system is an open research topic which has recently received renewed attention~\cite{kirkham2018defining}. 
In~\cite{milano2016frequency}, the point was raised that frequency is not uniform in the whole network, especially during transient conditions, and a formula to estimate the frequency at each system bus is proposed.
%
%Furthermore, in transient conditions, the frequency is not uniform in the whole network
%
In~\cite{freqcomplex}, the concept of \ac{cf} is introduced.
This complex variable is proposed as an extension to the well-known definition of frequency, i.e. as the time derivative of the argument of a sinusoidal signal~\cite{iec2018ieee}.
This extension quantifies the change of the network frequency caused not only by the variation of the phase angle, but also by the magnitude of the voltage. 
In this work, an analytical connection between the complex power injection and the complex frequency at each bus was described.
This connection is utilized to derive how devices that are found in traditional power systems, e.g., synchronous machines, affect the complex frequencies at the network buses.  
The developed methodology dropped the assumption of a lossless system and, therefore, presenting a generalised representation of the formula proposed in~\cite{milano2016frequency}.

The interpretation of \ac{cf} is approached in the literature from different points of view~\cite{paradoxes}.
The imaginary part of the frequency is, in effect, the conventional quantity that is commonly utilized, in signal processing and time-frequency analysis, to define the instantaneous frequency of a signal.
In the case of power systems, the signal is the voltage of (current injected into) a bus, and the instantaneous frequency is thus the time derivative of the phase angle of this voltage (current) when this is represented as an analytic signal.
Using the same signal processing approach, the real part of the frequency can be defined as instantaneous bandwidth~\cite{Cohen:1995}.
The geometric approach, presented in \cite{freqgeom}, assumes that the voltage (current) is the  speed of a trajectory in space.  This is supported by the fact that the voltage (current) is the time derivative of a time-varying flux (electric charge). This approach leads to the definition of the real and imaginary parts of the \ac{cf} as the symmetrical and anti-symmetrical components, respectively, of the time derivative of the voltage (current).
Finally, in \cite{freqfrenet}, based again on the analogy between voltage (current) and the speed of a trajectory, the real part of the \ac{cf} is interpreted as a radial speed, whereas the imaginary part is interpreted as an azimuthal speed.

The preliminary study on \ac{cf} (see reference~\cite{freqcomplex}) shows that such the formulation has relevant applications in modelling, control and state estimation.
Since its introduction, the concept of \ac{cf} has been used to develop a novel approach for power system state estimation~\cite{zhong2022line}.
In addition, it has been used to study the synchronization stability of converters using dispatchable virtual oscillator control~\cite{colombino2019global, he2022complex} as well as for finding equivalencies between the mentioned controller and a complex droop controller~\cite{he2023nonlinear}.
However, several other existing converter control schemes in the literature have not been studied previously under the lens of \ac{cf}.
Finally in~\cite{dervf}, the magnitude of the complex frequency was utilized as a metric to compare the performance of converter frequency and voltage controllers.
%In the same spirit, the present work is an application of the concept of complex frequency to the characterization of the dynamic behavior of the different parts that form the synchronization of the control of power electronics converters.

\vspace{-0.2cm}
\section{Main Contributions of the Thesis} \vspace{-0.2cm}

The main objective of the thesis has been to study the voltage stability of distribution networks with high penetration of power electronic converters and to propose new controllers with an aim to improve the voltage profiles in such networks.
A secondary objective has been to study the dynamic effect of different converter controllers on the network frequency accounting for both changes in the voltage magnitudes and phase angles.
The main contributions of the thesis are summarised in the following subsections.

\subsection{Bifurcation Analysis of Active Distribution Networks with High Penetration of Converter Interfaced Devices}
\begin{figure}
\centering
\includegraphics[width=1\columnwidth]{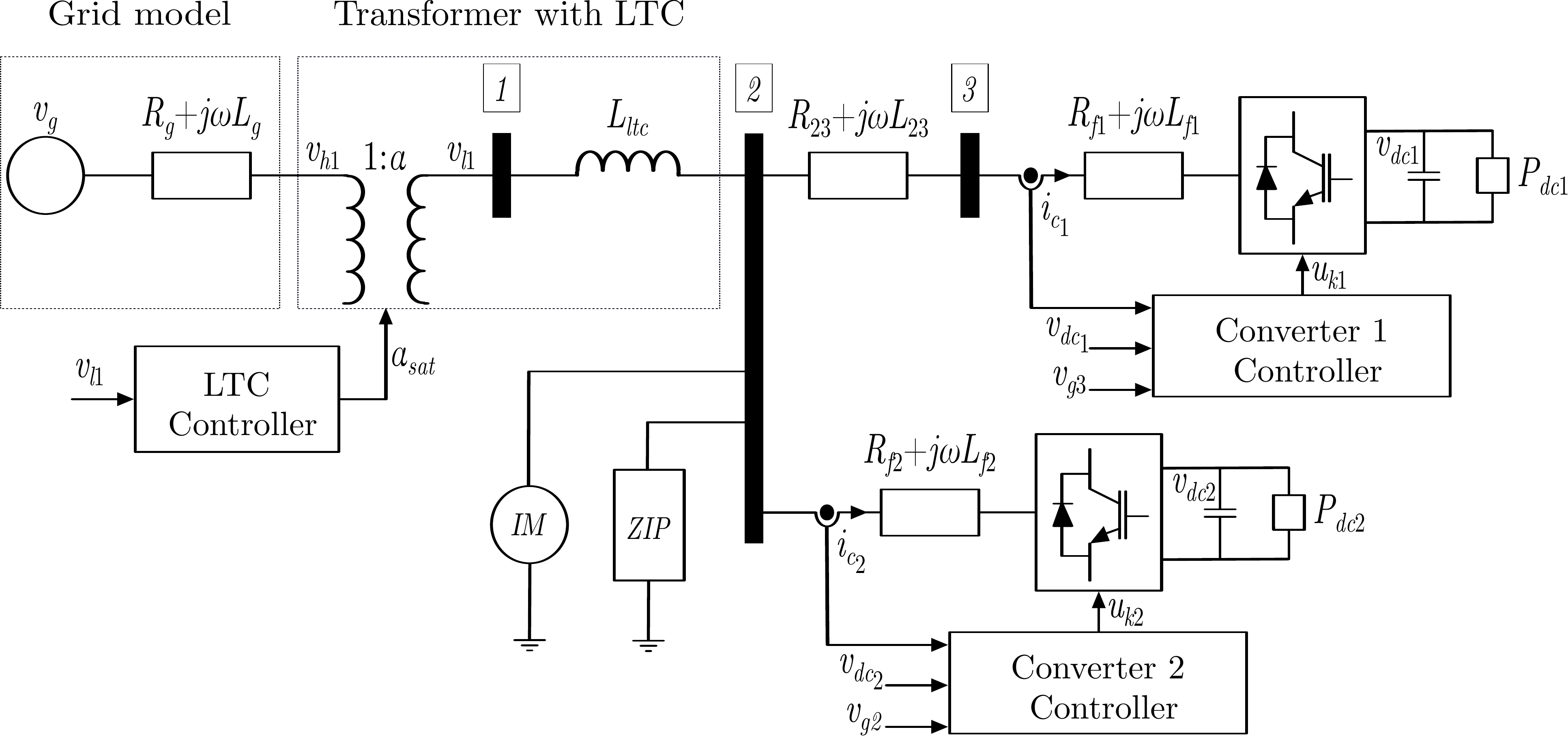}
\vspace{-0.1cm}
\caption{Representative distribution grid with high penetration of power converters that was used as a benchmark for the stability. It contains different devices that regulate voltage and affect voltage stability.}
\label{fig.grid_bifurc}
\end{figure}

In this work, stability boundaries of converter-dominated electrical distribution systems were investigated by using bifurcation analysis~\cite{moutevelis2022bifurcation}. 
Emphasis was given on voltage stability limits, related to the loadability of the system and on the converter control parameters that affect it.
%
%The full dynamic model of a representative distribution network was derived and parametric continuation was implemented. 
%
%Parametric continuation was implemented to first calculate the stability limits.
%
%Then, their type and characteristics were classified with the use of bifurcation theory.
%
The proposed analysis considers the nonlinear effects that the converter controllers introduce to the system.
This allows a proper assessment of the system operation, even for large deviations from the nominal conditions.

Figure~\ref{fig.grid_bifurc} shows a representative distribution network with high penetration of power converters that was used as a benchmark for the stability studies performed in this work.
Apart from the converters and their control systems, the network also includes a number of different devices that affect the voltage levels in a distribution network and its stability.
These devices are different type of loads, i.e. \acp{im} and \ac{zip} loads, as well as a transformer equipped with a \ac{ltc}.
Despite their relevant effect, these devices are rarely considered in power converter studies.
Their inclusion in this work, together with the study of their interaction with power converters, provides a comprehensive analysis of different factors that affect voltage stability in modern distribution networks.

\begin{figure}
\centering
\includegraphics[width=1\columnwidth]{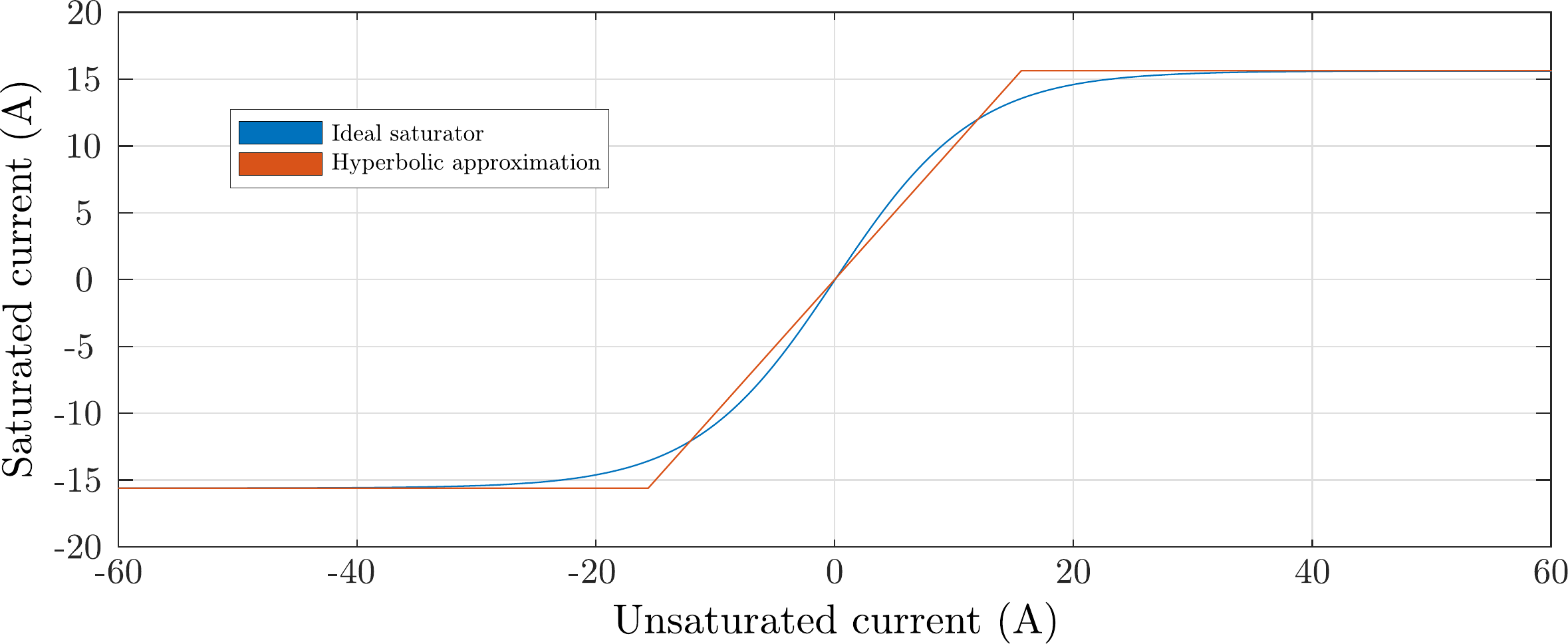}
\vspace{-0.1cm}
\caption{Approximation of an ideal saturation with a hyperbolic tangent
function. Application for the current saturation of a power converter based on its rated capacity.}
\label{fig.saturation_example}
\end{figure}

In this work, in order to include the saturation effects of the converters as well as the \acp{ltc} in the model, novel, smooth approximations were proposed.
These saturation effects, inherent to all devices with limited capacity, play an important and critical role to the stability of the system.
At the same time, they impose hard, non-smooth limits that may introduce numerical problems to the software used for the stability analysis.
Figure~\ref{fig.saturation_example} shows an example of the proposed approximations, based on hyperbolic functions.
The proposed formulation allows the incorporation of the saturation effects to the dynamic model of the system, avoiding the numerical issues caused by non-smooth limitations and without increasing the model complexity.

\subsection{Voltage control in distribution networks with distributed generation using the virtual admittance loop}

\begin{figure}[t]
\centering
\includegraphics[width=0.7\columnwidth]{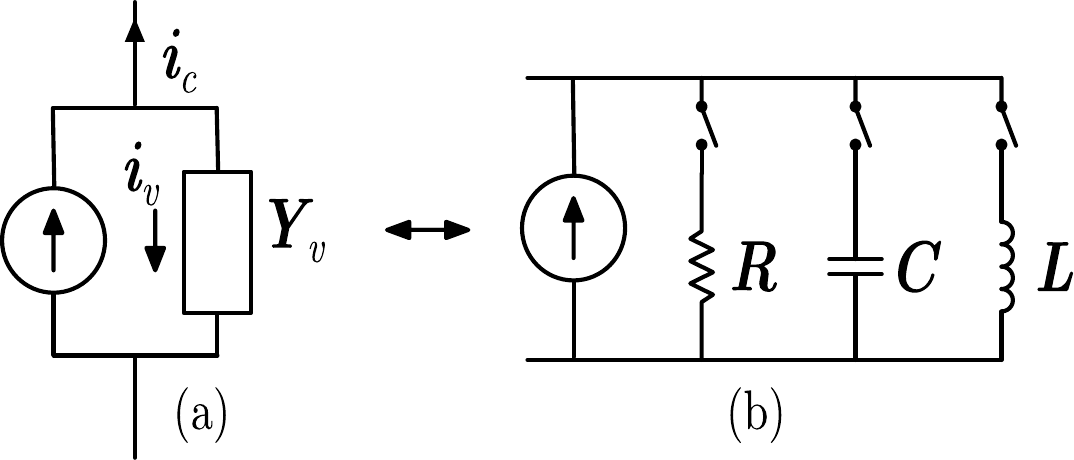}
\vspace{-0.1cm}
\caption{Schematic illustrating the equivalence of the proposed virtual admittance loop and a parallel, configurable admittance.}
\label{fig.val_equivalent}
\end{figure}
\begin{figure}
\centering
\includegraphics[width=1\columnwidth]{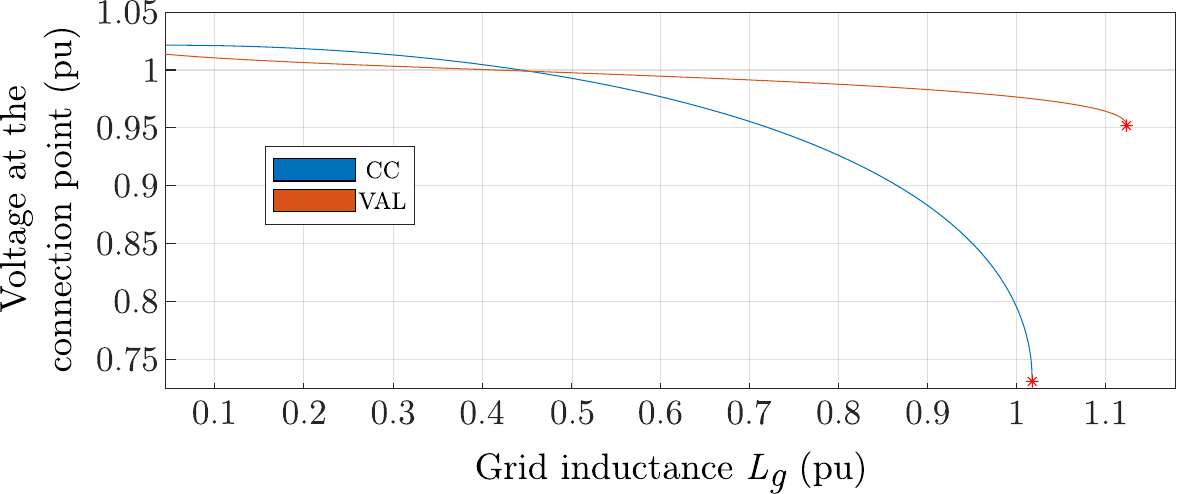}
\vspace{-0.1cm}
\caption{Voltage magnitude at the DG connection point for varying grid inductance. Comparison between a simple current controller and the proposed virtual admittance loop.}
\label{fig.val_example}
\end{figure}

In this work, a \ac{val} for \ac{dg} units was introduced with an aim to reduce voltage deviations and improve voltage stability. 
This control loop helps supporting the grid voltage using both active and reactive power, taking into consideration the natural coupling between them.
Compared to other combined active and reactive power voltage controllers in the literature, the \ac{val} does not treat this coupling as a side-effect, but instead utilises it to improve the effectiveness of the control. 
The proposed \ac{val} can be added on top of the typical, \ac{gfl} control structure comprised of a \ac{pll} and a \ac{cc}, facilitating its retrofit to already deployed converters. 
The work was divided in three parts.
First, the concept of virtual admittance was introduced and its application towards voltage regulation by \ac{dg} units was explained~\cite{moutevelis2021virtual}.
Then, an improvement upon the original implementation was proposed based on the quasi-stationary approximation of the derivative operator~\cite{moutevelis2022quasi}.
Finally, a recursive, secondary controller was developed for the optimal tuning of the different \acp{val} within the same distribution network~\cite{moutevelis2023recursive}.

Figure~\ref{fig.val_equivalent} illustrates the equivalency of the proposed \ac{val} with a set of configurable shunt conductances and susceptances, aimed at improving voltage levels.
The former is implemented inside the converter control, avoiding bulky and costly additional hardware, with the implemented control structure being equivalent with a parallel connection to the \ac{cc}, illustrated in the figure as an ideal current source.
The latter, especially shunt capacitors, have been typically used in the past for undervoltage support.
The motivation behind the \ac{val} is the realization that a \ac{dg} can operate seamlessly between inductive and capacitive modes while also adjusting its active power output.
%
%This capability resembles an operation of a variable admittance.
%
This operation extends the capabilities of traditional capacitor banks and can be leveraged to address both under- and overvoltages in resistive and inductive grids.
Figure~\ref{fig.val_example} shows the voltage magnitude at the \ac{dg} connection point for different grid inductance values.
It can be seen that, compared to the simple \ac{cc}, \ac{val} reduces the voltage deviation from the nominal point and improves the stability margin.

In its original implementation, \ac{val} emulated the full dynamics of an admittance in the controller.
For this reason it was also labeled as \ac{dval}.
In the following work, a new implementation for the \ac{val}, labeled as \ac{qval}, was proposed based on the quasi-stationary approximation.
This approximation assumes an emulation of the admittance dynamics only for the fundamental frequency~(50~Hz) and not for the full frequency range.
The \ac{qval} improves the transient performance of the controller and avoids instabilities caused by synchronous resonances.
At the same time it offers more flexibility for parameter tuning.
In this part, the quasi-stationary equations for the \ac{val} were firstly derived from the dynamic ones with the use of the quasi-stationary approximation of the derivative operator.
The theoretical formulations were complemented with Laplace domain dynamic analysis of the two control implementations and time domain simulations.
%

% %
\begin{figure}[t]
\centering
\includegraphics[width=0.8\columnwidth]{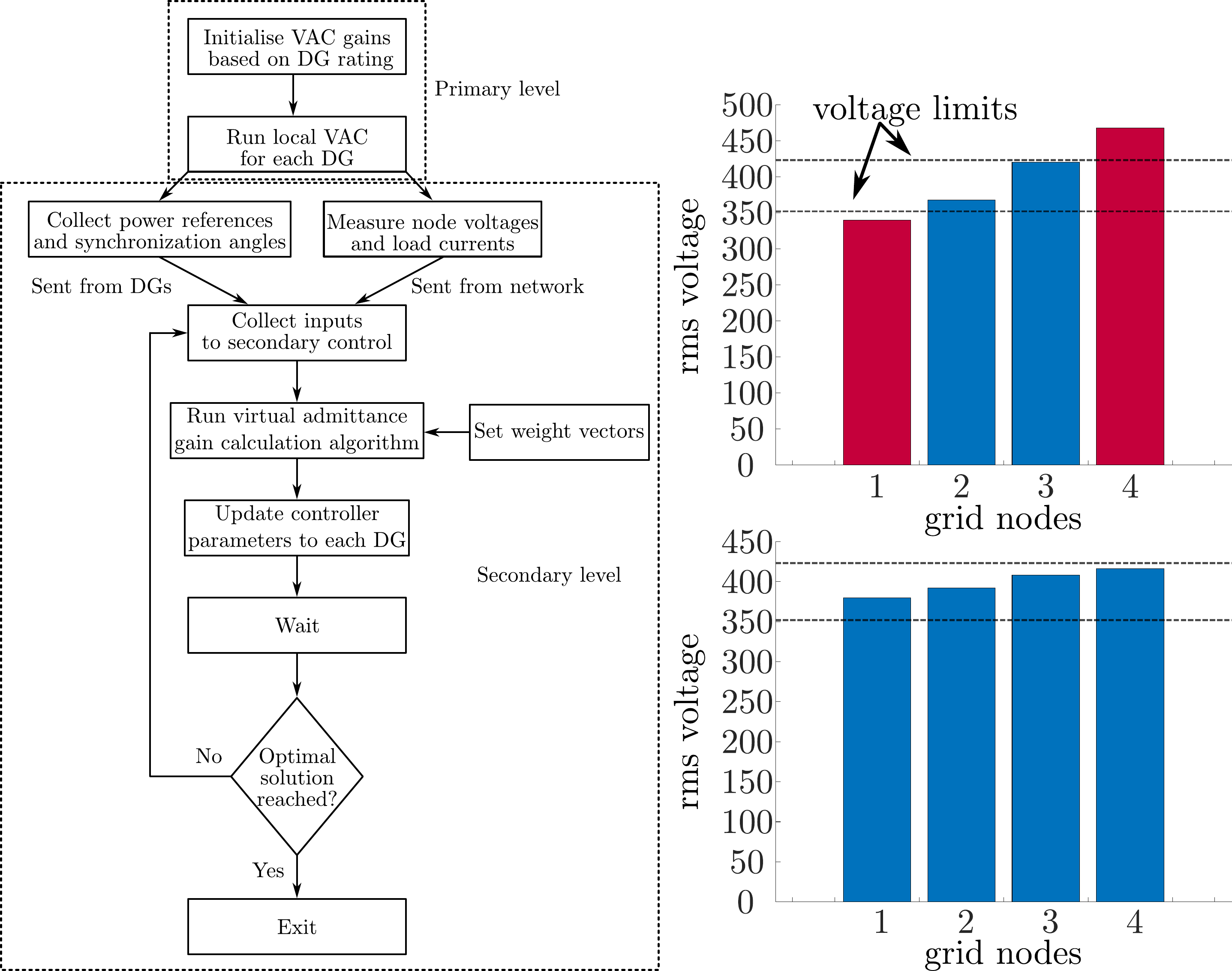}
\vspace{-0.1cm}
\caption{(left) Flow diagram with the steps of the proposed controller. 
(right) Example of rms voltages before and after applying the proposed controller.}
\label{fig.val_flowchart}
\end{figure}

The optimal tuning of all \ac{val} loops in the same distribution network was the last point to be addressed.
To this end, a recursive secondary controller was presented with an objective to improve the voltage profile in a distribution network with several \ac{dg} units.
The centralised control algorithm uses the measurements of the operating point from the grid to optimally tune \ac{val} parameters towards the goal of minimizing voltage deviations in the steady state and acts as a supplementary control layer to the local \acp{val}.
Figure~\ref{fig.val_flowchart} shows a flow diagram of the algorithm steps and an example showcasing its effect on the distribution grid voltage values. 
First, the local \acp{val} are enabled to control voltage only locally.
Then, the secondary controller is periodically enabled to further regulate the voltages in the distribution grid.
In order for the secondary controller to optimally tune the \acp{val} for all \ac{dg} units, the voltage measurements of each node and the current of each load need to be collected.
Converters also send to the secondary controller their active and reactive power set-points.
Then, the \ac{val} gains are calculated periodically and sent to the converters via a low-bandwidth communication channel.
By measuring network variables and sending them to the secondary control level, any change in the grid operating point is detected and taken into consideration by the algorithm in the following iteration.
This allows the secondary controller to optimize the voltage profiles based on the current state of the network.

The procedure described above provides a comprehensive solution for reducing voltage deviations in electrical distribution networks.
The solution is suitable for both over- and undervoltages as well as for networks with different $R/X$ ratios.
The proposed controllers address the voltage regulation both on the primary device level and on the secondary network level.
Both dynamic and steady state regulation aspects are taken into consideration.

\subsection{Taxonomy of power converter control schemes based on complex frequency}

\begin{figure}
\centering
\includegraphics[width=1\columnwidth]{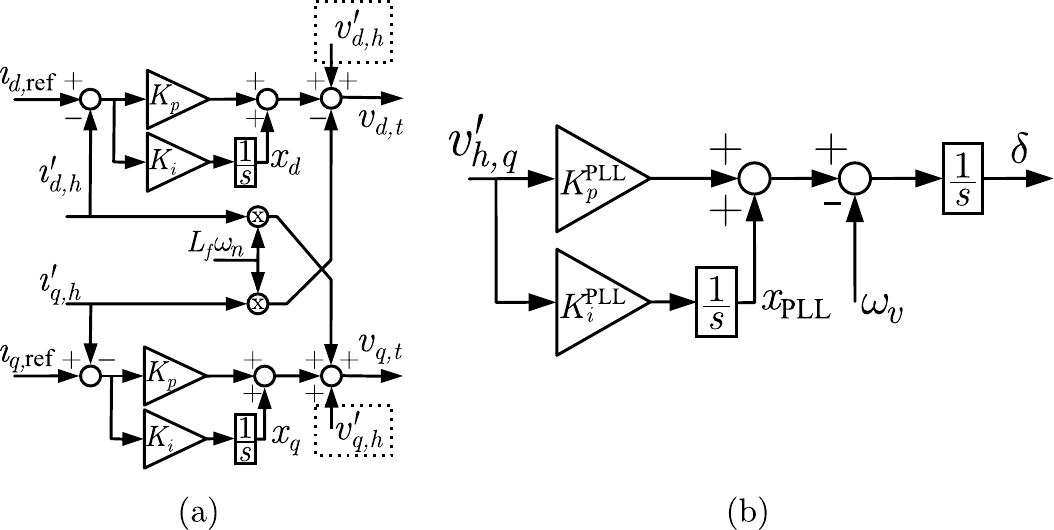}
\vspace{-0.1cm}
\caption{Examples of power converter control schemes whose impact on the local frequency is studied. (a) Current controller and (b) Phase-locked loop}
\label{fig.taxonomy_example}
\end{figure}

This work presented a systematic taxonomy of power converter synchronization and control schemes based on their effect on local frequency~\cite{moutevelis2023taxonomy}.
In modern power networks, the proliferation of converter-based \ac{dg} has led to the development of a range of different strategies for their control.
Figure~\ref{fig.taxonomy_example} shows two representative control schemes that are commonly used, i.e. the \ac{cc} and the \ac{pll}.
A common approach for the theoretical analysis of these control strategies consists in assuming they have sufficiently fast internal control loops so that they can be simplified and respresented only with unity gains.
This assumption, however, prevents fully understanding  the effect these internal controllers have on the network frequency. 
Furthermore, the increased penetration of these sources calls for their contribution to the frequency regulation in the network.
A number of such converter-based, frequency regulating strategies has been proposed in the literature with their stability and transient operation being well studied.
However, the contribution of each controller to the frequency at the converter connection point has not been fully discussed yet.

The recently proposed concept of \ac{cf} was applied in this work to formulate analytically the effects different power converter controllers have on the local frequency.
The \ac{cf} quantity was used to derive the effect of different converter control schemes on the grid frequency.
This novel approach allowed the decoupling of the effect of different participating controllers not only on the local frequency of the connection bus, but also on the transient voltage.
It also allowed the identification of critical control parameters. 
These parameters can be then tuned adequately to maximize or minimize the controller effect on the electric variables at the connection point.
The theoretical results were complimented by extensive sensitivity analysis and time domain simulations.

\section{Thesis Outline and Principal Results} 
The thesis is divided into three journal papers and two conference papers describing the fundamental advances.

%The thesis is divided into 7 chapters describing the fundamental advances and 3 appendices that include important content that can be omitted in a first read.

%Chapter 1 (this one) provides an introduction to the problems addressed in this work, presents motivation and objectives and summarizes the main contributions of the thesis. Chapter 2 introduces a full-state feedback control of Back-to-Back converters. Chapter 3 presents a control system for HVDC interconnections that provides damping of frequency oscillations in asynchronous AC grids. Chapter 4 introduces a small-signal modelling and stability analysis of VSCs connected to weak grids including the impact of the implementation effects. Chapter 5 presents a stability analysis of VSMs with virtual impedance in weak grids, and provides guidelines for the control parameter design. Chapter 6 presents a stability analysis of VSCs in MTDC networks for railway applications. Chapter 7 provides the conclusions of this work and proposes the guidelines for further research.

\subsubsection{Paper~1: ``Bifurcation Analysis of Active Electrical Distribution Networks Considering Load Tap Changers and Power Converter Capacity Limits''} 

This paper presents an analysis method to predict stability boundaries of converter-dominated electrical distribution systems by using bifurcation theory. 
To this end, a fully dynamic model of such a system is derived and parametric continuation was implemented by using dedicated software~\cite{dhooge2008matcont}. 
The analysis considered the nonlinear effects the converter controllers introduce to the system., including the often-ignored converter capacity limits.
Important voltage regulating devices like transformer \acp{ltc} were also considered.
Converter and \ac{ltc} limits were modelled by using smooth mathematical functions so that the complexity for resolving the problem did not increase.
This modelling strategy allows modelling of both static and dynamic saturation strategies.
The proposed analysis allows a proper assessment of the system operation, even when large deviations from the nominal parameters occur.
Stability properties and indices that cannot be derived by linearisation, for example limit cycle magnitudes or two-dimension stability boundaries, were provided.
The contribution of converters to the power system stability via reactive power support was thoroughly investigated.
This work shows that capacity limits of electrical distribution systems can be improved by adequately tuning the control parameters of electronic interfaces and that bifurcation theory is a useful tool for the selection of their values.
%

%This paper presents a full-state feedback controller for a back-to-back converter that provides a fast control of active and reactive powers and a stiff regulation of the DC-capacitor voltage. A power-based modelling approach based on the differential- and common-power (current) concepts are introduced. By using this modelling approach, the $LCL$ filter and DC-capacitor dynamics can be aggregated in a single state-space model. The proposed modelling approach allows the design of a single state-feedback controller based on the LQR theory that takes into account the dynamics of both the $LCL$ filters and the DC capacitor at the same time. The proposed controller has been used to achieve fast control of active and reactive powers and a stiff regulation of the DC-capacitor voltage at the same time. The robustness of this controller to variations of the grid inductance, the filter parameters and the DC capacitor values are studied. The analysis revealed that controller is robust against variations on this parameters. Also, a current reference generation and limiting system is proposed and a detailed analysis of the DC capacitor dynamics and power stored in the connection filters are presented.

This paper was published in the journal \emph{IEEE Transactions on Power Electronics}~\cite{moutevelis2022bifurcation} and presented at the IEEE Energy Conversion Congress and Exposition (ECCE 2020)~\cite{moutevelis2020bifurcation}.

\subsubsection{Conference~1: ``Virtual Admittance Control for Providing Voltage Support using Converter Interfaced Generation''}

This paper proposes a control loop for \ac{dg} units that aims to improve the voltage profile of electrical distribution networks. 
This controller is an outer, circuit-inspired loop that is based on the dynamic operation of an electrical admittance.
The aim of this control loop is to support the grid voltage with both active and reactive power, taking into consideration the natural coupling between them in highly resistive networks.
This control strategy takes into account the restrictions of active and reactive power imposed by the \ac{dg} hardware and the primary source of energy and can be implemented with minimal alterations to existing control schemes.
The performance of the proposed controller in improving the voltage profile was showcased by using continuation analysis and time domain simulations.
%

%This paper presents a small-signal modelling and stability analysis of VSCs in multi-terminal DC (MTDC) networks with energy storage system for railway applications. The VSCs were modelled independently, including the dynamics properties of output filters and control systems, while the common DC-busbar was modelled as an equivalent capacitor that aggregates the total capacitance of the DC-link. These models were linearised and joined together to form a single small-signal state-space model, including the interactions between VSCs thought the common DC-link. The stability of the MTDC system is studied by using eigenvalue analysis. The impact of the droop gains on the stability of the MTDC system is studied by using the eigenvalue analysis, revealing the theoretical limits of the stable operation of the system. 

This paper was presented at the 2021 IEEE PES Innovative Smart Grid Technologies Europe (ISGT Europe)~\cite{moutevelis2021virtual}.

\subsubsection{Conference~2: ``Quasi-Stationary Implementation of Virtual Admittance Controller for
Voltage Support from Distributed Generation''}

This paper proposes an improvement of the controller presented in Conference~1 based on a quasi-stationary approximation of the admittance dynamics inside the controller.
The resulting implementation improves the transient performance of the controller and avoids instabilities triggered by synchronous resonances.
At the same time, more flexibility for parameter tuning is provided.
In this work, the derivation of the quasi-stationary equations from the dynamic ones was explained.
Laplace domain dynamic analysis was performed for the two implementations to assess their dynamic operation and to compare them directly.
The analytical results provided by the dynamic analysis were validated by using time domain simulations.
%

%This paper presents a small-signal modelling and stability analysis of VSCs in multi-terminal DC (MTDC) networks with energy storage system for railway applications. The VSCs were modelled independently, including the dynamics properties of output filters and control systems, while the common DC-busbar was modelled as an equivalent capacitor that aggregates the total capacitance of the DC-link. These models were linearised and joined together to form a single small-signal state-space model, including the interactions between VSCs thought the common DC-link. The stability of the MTDC system is studied by using eigenvalue analysis. The impact of the droop gains on the stability of the MTDC system is studied by using the eigenvalue analysis, revealing the theoretical limits of the stable operation of the system. 

This paper was presented at the 2022 IEEE PES Innovative Smart Grid Technologies Europe (ISGT Europe)~\cite{moutevelis2022quasi}.

\subsubsection{Paper~2: ``Recursive Secondary Controller for Voltage Profile Improvement Based on Primary Virtual Admittance Control''}

This paper presents a recursive, secondary controller that is based on the virtual admittance concept and aims to improve the voltage profile of a distribution network in which several \acp{dg} coexist.
This represents an extension of the work originally presented in Conferences~1 and~2, in which only the primary virtual admittance controller was considered.
The proposed method addresses two problems that are often considered separately in the literature, namely both under- and overvoltage conditions.
The centralised algorithm uses measurements of the operating point of the grid to optimally tune the converter controllers.
Acting as a supplementary control, the method does not require any high bandwidth communication links.
The effectiveness of the proposed controller was showcased by using time domain simulations and experimental tests.
%

%This paper presents a control scheme for HVDC interconnections that provides oscillation damping capabilities in asynchronous AC grids. The HVDC link is modelled by using the differential- and common-power concepts, which allows an independent control of the DC-link voltage and the net power transfer between AC grids. The proposed controller introduces a virtual electro-mechanical effect which is equivalent to a mechanical friction coupling the inertial dynamics of the two AC grids. This virtual friction controller provides frequency support as well as POD of area oscillation modes (intra-area oscillations) at both sides of the HVDC link without relying on fast communications between the link terminals. The impact of the virtual friction effect on the stability is analysed analytically by using a simplified model composed of two rotating masses coupled by a mechanical friction. Moreover, the impact of the local oscillation modes of the AC grid on the POD capability of the virtual friction controller is evaluated as well as the impact of the DC-line resistance.

This paper was published in the journal \emph{IEEE Transactions on Smart Grid}~\cite{moutevelis2023recursive}.

\subsubsection{Paper~3: ``Taxonomy of Power Converter Control Schemes based on the Complex Frequency Concept''}

This paper proposed a framework to calculate the impact of different converter control structures on the bus frequency at the point of converter connection.
To this end, the mathematical properties of the \ac{cf} quantity were combined with the dynamic equations of different controllers, resulting in novel mathematical formulations for each controller type.
These formulations provide analytical insight of the effect of each sub-controller and control parameter to the frequency of the voltage at the converter bus, as is explained below.
%
%mathematical formulations utilising the novel concept of \ac{cf} and combining it with the dynamic equations for different controllers were produced. 
%

The notion of voltage \ac{cf} had been previously presented in the literature to unify in a single complex quantity the effect of variations of the voltage phase angle and magnitude on frequency.
In this paper, the notion was extended to other variables and signals like currents and voltage references.
It was explained how \ac{cf} can be used as a derivative operator for such signals and how the generalised \ac{cf} quantity, resulting from the above extension, can be used to derive the effect of different converter control schemes on the grid frequency.
This novel approach allows the decoupling of the effect of different participating controllers, often present in complex cascaded configurations, as well as the identification of critical control parameters.
These parameters can be then tuned adequately to change the controller effect on the grid frequency.
The local frequency as perceived by the converter was calculated and categorized.
This internal frequency differs from the bus frequency due to the action of the converter controllers and synchronization mechanism.
A parallel with the rotor speed of a synchronous machine was drawn, allowing the study of synchronous and asynchronous generation types using the same theoretical tools.
Applications of the internal frequency for the improvement of the converter control operation were showcased.
The theoretical contributions of this work were supported by time domain simulations considering various controller configurations, both \ac{gfl} and \ac{gfm}, using a realistic grid benchmark.
%

%This chapter presents a detailed small-signal model of voltage source converters connected to weak grids. The proposed modelling approach includes effects such as dead-time and computational time-delays caused by the digital implementation and the switching process. The impact of the dead-time effect on the converter small-signal model is evaluated for different operating points, revealing that it produces a damping effect related to the output current level delivered by the converter. The analysis of the time-delay revealed that it has a significant impact on the converter stability and should be included in the small-signal model, especially if $LCL$ filters are used. In this chapter, it will be shown that the Singular Value Decomposition (SVD) analysis can effectively predict harmonic oscillations and frequency ranges in which inputs are prone to produce network resonances.

This paper has been accepted for publication in the journal \emph{IEEE Transactions on Power Systems}~\cite{moutevelis2023taxonomy}.

\chapter{Conclusions and Suggestions for Further Research} 
\label{cap.conclusion}

% Cabecera
\pagestyle{fancy}
\fancyhf{}
\fancyhead[LE,RO]{\thepage}
\fancyhead[RE,LO]{Chapter \thechapter. Conclusions and Further Research}

The objective of this thesis is to propose new methods for evaluating voltage stability in modern, electrical distribution networks and to propose novel ways of leveraging the control capabilities of power converters, safeguarding the operation of such networks with high penetration of renewable energy sources.
In order to achieve this objective, this thesis addressed multiple aspects of voltage stability.
A stability analysis method based on bifurcation theory was applied considering critical and often ignored effects, like converter capacity limits and \acp{ltc}.
The proposed approach can serve as a framework for assessing voltage and parameter stability in any converter-dominated electrical distribution network.
A complete control solution, implemented on both primary and secondary control levels, was proposed for power converters in distribution networks with voltage deviation issues.
The proposed controllers were based on the virtual admittance concept and combine the advantages of local and network-wide control without the requirement of fast communication infrastructure, addressing under- and over-voltages, simultaneously.
Thanks to the use of complex frequency, the problem of transient voltage instability, caused by power electronic converters, was studied together with frequency instability.
Detailed theoretical formulations were developed to explain the effect of different configurations of power converter controllers on the local voltage and frequency transient operation.
By doing so, the proposed framework provides insight for the optimal tuning of existing voltage controllers and paves the way for the development of new ones.
\section{Main Contributions}
The original contributions of the thesis are summarised as follows:

\begin{itemize}

\item Application of \textbf{bifurcation theory} has been proposed as a novel approach for the \textbf{stability analysis of distribution networks with power converters}. Compared to standard methods, the theory expands the parameter region of study, allowing \textbf{large deviations} from the nominal values. In addition, it accounts for \textbf{nonlinear effects} in such networks, something that cannot be achieved by using linearisation-based techniques.

\item A novel modelling technique has been proposed to include \textbf{\acp{ltc}} and \textbf{converter capacity limits} to \textbf{bifurcation studies}. The approach is based on approximating non-smooth saturation functions with \textbf{smooth function equivalents}. Both \textbf{static} and \textbf{dynamic saturation limits} are taken in consideration. The modelling technique was validated through comprehensive comparisons with realistic, non-smooth saturation functions that are often used in commercial converter control schemes.

\item A circuit-inspired, \textbf{virtual admittance} based controller has been proposed for \acp{dg} units to \textbf{regulate voltage in resistive and weak electrical distribution networks}. The proposed controller changes the active and reactive power set-points of the converter simultaneously, taking advantage of the natural coupling between the two in resistive grids. In addition, an improved implementation of the above controller is also proposed. This implementation is based on a \textbf{quasi-stationary approximation} of the virtual admittance dynamics and improves the overall performance of the controller.

\item A \textbf{recursive, secondary voltage controller} has been proposed that acts upon the proposed local, virtual admittance controller. The secondary controller periodically receives measurements from the grid and the \acp{dg} and updates the control parameters of the local controllers by using an optimisation algorithm. The optimisation objective is to minimize the voltage deviation from the nominal set-point across the grid so that both \textbf{under-} and \textbf{over-voltage} cases are addressed. \textbf{Network topology} and \textbf{converter current restrictions} are included in the algorithm as constraints while \textbf{weight vectors} are included in the objective function to provide scheduling flexibility.

\item A \textbf{taxonomy of different power converter control schemes} has been proposed based on their impact on the frequency at their connection point. The taxonomy was developed by using the recently proposed quantity of \textbf{\ac{cf}} that is an extension of the well-known frequency definition as the derivative of the voltage phase angle. The proposed approach allows the analytic study of the specific contribution to frequency of different controller, including \ac{gfl} and \ac{gfm} control structures.

\end{itemize}

\section{Conclusions}
The main conclusions obtained during the development of this thesis are summarised as below:

\begin{itemize}

\item Bifurcation analysis was shown to be an effective tool for studying parameter stability limits in distribution systems with a high penetration of converter interfaced loads and generators. The method allows the stability limits to be sought under large parameter variations while assuming these variations occur slowly. In that regard, it complements standard small-signal analysis methods that study small deviations from the network nominal operation point and Lyapunov-based methods that study sudden, large transient events.

\item The smooth mathematical functions that are proposed in Paper~1 were able to approximate with high accuracy the typically non-smooth saturation functions, present in industrial converters and transformers with \acp{ltc}. This allowed the incorporation of such devices in a comprehensive model that is suitable for bifurcation analysis studies.

\item  Non-linear effects like converter saturation was shown to cause system instability under seemingly safe operation conditions. Voltage regulating control actions, like converter reactive power support and transformer \ac{ltc} regulation, were shown to improve the parameter stability margins of the system.

\item Through bifurcation analysis, useful stability properties and indices were calculated. These include the magnitudes of encountered limit cycles and two-dimension parameter stability diagrams. The latter were proven particularly useful for the selection of converter control parameters.

\item The virtual admittance controller for \acp{dg} of Conference~1 was shown to successfully provide voltage support in weak, resistive distribution networks. When compared to a standard, \ac{gfl} converter with a current controller, it was shown to reduce deviations from the nominal voltage and improve the system stability margins. Under stochastic grid voltage profile variations in a resistive network, it provided improved voltage regulation at the converter connection point compared to a standard voltage-current droop controller. 

\item The quasi-stationary implementation of the virtual admittance controller of Conference~2 avoids the resonant instabilities of the dynamic implementation and improves the overall dynamic performance. At the same time, it allows wider control parameter selection range.

\item The secondary controller based on virtual admittance further improves the voltage regulation in distribution networks compared to the application of the primary controller only. Both over- and undervoltages are addressed.

\item  The participation of each \ac{dg} is properly adjusted by the secondary controller weight selection, allowing additional flexibility to the network operator.

\item The combined primary and secondary virtual admittance control setup outperforms common active/reactive power droops over a large range of system parameters and operating points.

\item \ac{cf} was shown to be an effective theoretical tool for the decoupling the contributions of each sub-controller in complex control configurations  to the local frequency  as well as for the identification of critical control parameters.

\item  The proposed theoretical framework allows the calculation of the internal frequencies for different control topologies. These internal frequencies were shown to be of practical interest for the improvement of the converter frequency response compared to the use of exact, local signals. The study case examples showed that the use of the internal frequencies, under some specific conditions, can replace the exact, local signals (in case they are unavailable) and may even improve the dynamic performance of the control.

\item The use of \ac{cf} allowed the joint study of voltage and frequency stability in converter-dominated power systems with the use of the same theoretical tools.

\end{itemize}

\section{Further Research Suggestions}
The results obtained in this thesis create the following research opportunities:
\begin{enumerate}

\item Paper~1 applied bifurcation theory to analyse the stability of a distribution network with high penetration of power converters. Apart from their radial topology that was addressed, other characteristics of such networks is the presence of unbalanced loads and stochastic load variations. Future research could focus on applying the proposed method to unbalanced and stochastic systems. The application of this method in larger and of meshed topology electrical distribution grids is also of interest. 

\item In Paper~1, only the impact of \ac{gfl} converters on the system stability was studied. Also, they were only equipped with standard reactive power compensators for the voltage regulation. A larger variety of converter control structures, including \ac{gfm} topologies, could be studied in future research.

\item The proposed primary and secondary virtual admittance controllers, proposed in Conferences~1 and~2 and in Paper~2, do not consider the primary source of the \ac{dg} units. It would be of interest to study the impact of the primary source variability on the controller performance. Furthermore, the study of the dynamic operation of different sources like batteries, photovoltaics or wind turbines and the restrictions they impose on the controllers would be of merit.

\item In Paper~2, the effectiveness and performance of the proposed secondary controller was demonstrated via extensive simulations and laboratory experiments. However, a theoretical study of its stability under different operation scenarios is missing. This could serve as an interesting topic for future research.

\item The secondary controller in Paper~2 was designed for applications in resistive, electrical distribution networks. However, its adaptation for voltage regulation in high-voltage, meshed transmission networks would be of interest. In particular, the effect of the controller in inductive grids, its interaction with other voltage regulating devices~(e.g. STATCOMS) and its computational scalability in larger networks are all topics worth studying. 

\item The application of \ac{cf} in power control applications is a promising topic for further research. More specifically, the internal frequencies of the converters can be used as alternative inputs to standard auxiliary controllers, e.g. for the damping of low frequency oscillations. The use of non-standard control signals, for example the real part and the magnitude of the complex frequency, can be also explored. Finally, the potential of using the complex formulations of Paper~3 to develop novel, non-conventional controllers can be investigated.

\item The effect on \ac{cf} of multiple converters and their dynamic interaction can be further studied. In particular, it would be of interest to develop new metrics based on \ac{cf} that can capture the propagation of oscillations in networks with high penetration of power converters.

\end{enumerate}

\section{Publications}
\subsection{Journal Papers}
Research work included in this thesis has given rise to three JCR-indexed journal papers:

\begin{enumerate}

%\item D. Moutevelis, J. Rold\'{a}n-P\'{e}rez, M. Prodanovic and F. Milano, ``Taxonomy of Power Converter Control Schemes based on the Complex Frequency Concept,'' in \textit{IEEE Transactions on Power Systems} (early access).

\item D. Moutevelis, J. Rold\'{a}n-P\'{e}rez, M. Prodanovic and F. Milano, "Taxonomy of Power Converter Control Schemes Based on the Complex Frequency Concept," in \textit{IEEE Transactions on Power Systems}, vol. 39, no. 1, pp. 1996-2009, Jan. 2024.

\item D. Moutevelis, J. Rold\'{a}n-P\'{e}rez, N. Jankovic and M. Prodanovic, "Recursive Secondary Controller for Voltage Profile Improvement Based on Primary Virtual Admittance Control," in \textit{IEEE Transactions on Smart Grid}, vol. 14, no. 6, pp. 4296-4311, Nov. 2023.
%, doi: 10.1109/TSG.2023.3252803.

%D. Moutevelis, J. Rold\'{a}n-P\'{e}rez, N. Jankovic and M. Prodanovic, ``Recursive Secondary Controller for Voltage Profile Improvement Based on Primary Virtual Admittance Control,'' in \textit{IEEE Transactions on Smart Grid} (early access).

\item D. Moutevelis, J. Rold\'{a}n-P\'{e}rez, M. Prodanovic and S. Sanchez-Acevedo, ``Bifurcation Analysis of Active Electrical Distribution Networks Considering Load Tap Changers and Power Converter Capacity Limits,'' in \textit{IEEE Transactions on Power Electronics}, vol. 37, no. 6, pp. 7230-7246, June 2022.

\end{enumerate}

\subsection{Conference Papers}
Research work included in this dissertation has been presented in international conferences, giving rise to two conference papers:
\begin{enumerate}

\item D. Moutevelis, F. Göthner, J. Rold\'{a}n-P\'{e}rez and M. Prodanović, ``Quasi-Stationary Implementation of Virtual Admittance Controller for Voltage Support from Distributed Generation," \textit{2022 IEEE PES Innovative Smart Grid Technologies Conference Europe (ISGT-Europe)}, 2022, pp. 1-5.

\item D. Moutevelis, J. Rold\'{a}n-P\'{e}rez and M. Prodanovic, ``Virtual Admittance Control for Providing Voltage Support using Converter Interfaced Generation," \textit{2021 IEEE PES Innovative Smart Grid Technologies Europe (ISGT Europe)}, 2021, pp. 01-06.

%\newpage
%\vspace{0.5cm}
\setlength{\parindent}{-1cm}{
In additional, conference papers related to the thesis that were not included:
}

\item D. Moutevelis, J. Rold\'{a}n-P\'{e}rez, M. Prodanovic and F. Milano, ``Design of Virtual Impedance Control Loop using the Complex Frequency Approach'', in \textit{IEEE Powertech 2023 Conference}, Belgrade, Serbia.

\item D. Moutevelis, J. Rold\'{a}n-P\'{e}rez and M. Prodanovic, ``Aggregate Model of Parallel Distributed Energy Resources Controlled using Virtual Admittance'', in \textit{IEEE CPE-Powereng 2023 Conference}, Tallinn, Estonia.

\item J. Rold\'{a}n-P\'{e}rez, D. P. Mor\'{a}n-Rio, D. Moutevelis, P. Rodr\'{\i}guez-Ortega, N. Jankovic, M. E. Zarei and M. Prodanovic, ``Emulation of Complex Grid Scenarios by using Power Hardware In the Loop (PHIL) Techniques,'' in \textit{IECON 2021 – 47th Annual Conference of the IEEE Industrial Electronics Society}, 2021, pp. 1-6.

\item D. Moutevelis, J. Roldán-Pérez, M. Prodanovic and S. S. Acevedo, ``Bifurcation Analysis of Converter-Dominated Electrical Distribution Systems,''in \textit{2020 IEEE Energy Conversion Congress and Exposition (ECCE)}, Detroit, MI, USA, 2020, pp. 1670-1677.

\end{enumerate}

	\cleardoublepage
\addcontentsline{toc}{chapter}{Bibliography}
\bibliographystyle{IEEEtran}

% Cabecera
\pagestyle{fancy}
\fancyhf{}
\fancyhead[LE,RO]{\thepage}
\fancyhead[RE,LO]{Bibliography}
\renewcommand{\footrulewidth}{0pt}

\bibliography{My_Collection_Tesis}
	%\thispagestyle{empty}
	%\thispagestyle{empty}
	%\cleardoublepage
%\appendix
	%\include{AP_Capacitor_Linear}
	%\include{AP_BTB_Matrices}
	%\include{AP_PowerSystemModel}
\end{document}